\documentclass{aa}
\usepackage{graphicx}
\usepackage{natbib}
\usepackage{epstopdf}
\bibpunct{(}{)}{;}{a}{}{,}
\usepackage{txfonts}
\usepackage{longtable}

\begin{document}
\title{Detailed abundance analysis from integrated high-dispersion spectroscopy: Globular clusters in the Fornax Dwarf Spheroidal\thanks{
Based on observations made with ESO Telescopes at the La Silla Paranal Observatory under programme ID 078.B-0631(A)
}
}

\author{S. S. Larsen \inst{1} \and
  J. P. Brodie \inst{2} \and
  J. Strader \inst{3,4}
}
\institute{Department of Astrophysics / IMAPP, Radboud University Nijmegen, P.O. Box 9010, 6500 GL Nijmegen, The Netherlands
\and
  UCO/Lick Observatory, University of California, Santa Cruz, CA 95064, USA
\and
  Harvard-Smithsonian Center for Astrophysics, 60 Garden Street, Cambridge, MA 02138, USA
\and
  Department of Physics and Astronomy, Michigan State University, East Lansing, Michigan 48824, USA
}

\offprints{S.\ S.\ Larsen, \email{s.larsen@astro.ru.nl}}

\date{Submitted  26 June 2012 / Accepted 3 September 2012}

\abstract
{}
{
We describe our newly developed approach to detailed abundance analysis from integrated-light high-dispersion spectra of star clusters.  As a pilot project, we measure abundances of several Fe-peak, $\alpha$- and neutron capture elements from spectra of three globular clusters (GCs) in the Fornax dwarf spheroidal galaxy, obtained with UVES on the ESO \textit{Very Large Telescope}.
}
{
We divide the cluster colour-magnitude diagrams  into about 100 bins and use the Kurucz \texttt{ATLAS9} and \texttt{SYNTHE} codes to compute synthetic spectra for each bin. Stellar parameters are derived empirically from \textit{Hubble Space Telescope} data for the brighter stars, while theoretical isochrones are used for extrapolation below the detection limit. The individual model spectra are co-added and the abundances are iteratively adjusted until the best match to the observed spectra is achieved.
}
{
We find [Fe/H] = $-2.3$, $-1.4$ and $-2.1$ for Fornax 3, 4 and 5, 
with estimated $\pm0.1$ dex uncertainties. Fornax 3 and 5 are thus similar in metallicity to the most metal-poor Milky Way GCs and fall near the extreme metal-poor end of the field star metallicity distribution in Fornax. The [$\alpha$/Fe] ratios, as traced by Ca and Ti, are enhanced with respect to the Solar composition at the level of $\sim+0.25$ dex for Fornax 3 and 5, and possibly slightly less ($\sim+0.12$ dex) for Fornax 4. For all three clusters the $[$Mg/Fe$]$ ratio is significantly less elevated than [Ca/Fe] and [Ti/Fe], possibly an effect of the abundance anomalies that are well-known in Galactic GCs. We thus confirm that Mg may be a poor proxy for the overall $\alpha$-element abundances for GCs.
The abundance patterns of heavy elements (Y, Ba and Eu) indicate a dominant contribution to nucleosynthesis from the $r$-process in all three clusters, with a mean [Ba/Eu]$\sim-0.7$, suggesting rapid formation of the GCs.
}
{
Combining our results with literature data for Fornax 1 and 2,  it is now clear that four of the five Fornax GCs fall in the range $-2.5<{\rm [Fe/H]}<-2$, while Fornax 4 is unambiguously and substantially more metal-rich than the others. The indications that abundance anomalies are detectable in integrated light are encouraging, particularly for the prospects of detecting such anomalies in young, massive star clusters of which few are close enough for individual stars to be observed in detail.
}


\titlerunning{High-dispersion spectroscopy of the Fornax dSph GCs}
\maketitle

\section{Introduction}

The chemical composition of stellar populations as a function of time and location holds vital clues to the evolution of their host galaxies. The distribution of  metallicities for G-dwarfs in the Galactic disc implies significant gas infall over the lifetime of the disc \citep{VandenBergh1962,Pagel1975,Kotoneva2002,Naab2006}, and age-metallicity and Galactocentric distance-metallicity trends   provide additional important constraints on models for Galactic chemical evolution \citep{Matteucci1989,Pagel1995,Chiappini1997,Chiappini2001,Marcon-Uchida2010}. An even more detailed picture comes from the abundance patterns of individual elements, which are sensitive to the time scales of star formation. Perhaps the best-known example is the ratio of $\alpha$-capture to Fe-peak elements, which is an indicator of the relative contributions to nucleosynthesis by (short-lived) core-collapse supernovae vs.\ type Ia SNe \citep{Tinsley1979,Matteucci1986}. The ancient stellar populations in the Milky Way halo and bulge, as well as giant early-type galaxies, are typically enhanced in the $\alpha$-elements, indicating that star formation was a relatively rapid process \citep{Worthey1998,Trager2000,Thomas2005}. Further information about time scales and nucleosynthetic sites comes from the abundances of neutron-capture elements \citep{McWilliam1997} and even within the Fe-peak group, different elements display significant abundance variations that may hold important clues to the detailed nucleosynthetic histories of stellar populations.

Although a vast body of work exists on the chemical composition and evolution of both Galactic and extragalactic stellar populations, there is  a striking difference between the level of detail that can typically be achieved in these two domains. In the Milky Way, high-dispersion spectroscopy of individual stars provides a highly detailed picture of the abundance patterns of individual elements \citep{Edvardsson1993,Reddy2006}, either via measurements of the equivalent widths of individual lines or, when line blending gets too severe, spectral synthesis. Such measurements can be made for stars of essentially all ages, although the most detailed measurements remain constrained to the relative vicinity of the Sun \cite[though recently pushed to dwarf- and subgiant stars in the bulge via microlensing;][]{Bensby2011}.
For external galaxies, constraints on chemical composition must typically  rely either on measurements of emission lines in \ion{H}{ii} regions (for star-forming galaxies) or on relatively low-dispersion spectroscopy of integrated stellar light. In most cases,  interstellar gas abundances are only available for a limited number of elements (mainly O, N, S) and rely on empirical and uncertain calibrations of line strength vs.\ abundance \citep{Stasinska2010}. Furthermore, emission lines only provide a snapshot of the present-day \emph{gas} chemistry and may be biased by effects such as segregation onto dust grains \citep{Peimbert2010}. 

Integrated-light measurements of stellar absorption features are typically carried out by means of pre-defined sets of absorption-line indices, the best-known being the Lick/IDS system \citep{Worthey1994}. These indices are relatively broad ($\sim10-50$\AA ) and well-suited for measuring strong absorption features in the integrated light of early-type galaxies, where the spectral resolution is in any case limited by the internal velocity dispersions of the galaxies. Although the various indices have different sensitivities to different elements (such as Fe, Mg and the hydrogen Balmer lines),  none of them cleanly measures individual absorption lines. Nevertheless, the Lick/IDS system has been very successful in providing constraints on (luminosity-weighted mean) ages and overall metallicities of extragalactic stellar populations, and even on some abundance patterns such as overall [$\alpha$/Fe] ratios \citep{Kuntschner2000,Trager2000,Puzia2005,Thomas2005}.  
Physical properties such as ages and metallicities are normally inferred from such data by comparison with predictions by ``simple stellar population'' (SSP) models, either by simply plotting suitable combinations of index measurements on top of a model grid  or in a more automated fashion \citep[e.g.][]{Cenarro2007,Graves2008,Proctor2004,Strader2004}. In some cases, more sophisticated inversion techniques have also been applied to infer star formation histories from spectroscopic data \citep{Heavens2004,CidFernandes2005,Walcher2010}.

Unlike the spectra of giant elliptical galaxies, the spectra of most star clusters are velocity broadened by only a few km s$^{-1}$, allowing observations at much higher spectral resolution.
With efficient high-dispersion spectrographs on 8--10 m class telescopes, such as VLT/UVES and Keck/HIRES,  high S/N spectra can be obtained for star clusters well beyond the Local Group. 
Such observations have the potential to provide a vastly more detailed picture of chemical abundance patterns compared to techniques developed for integrated-light spectra of galaxies. Furthermore, star clusters provide discrete sampling points in time and space, even for stellar populations that contribute only a small fraction of the integrated light (such as metal-poor stellar halos).
However, the tools required to interpret such data have, until recently, not been available. 

Over the past decade, our group has been active in collecting high-dispersion spectra of extragalactic star clusters. These have been used primarily for dynamical mass measurements \citep{Larsen2002b,Larsen2004a,Larsen2004b,Strader2009,Strader2011}, but in many cases the S/N is also adequate for abundance analysis. With the next generation of extremely large telescopes, such data will be within reach at the distance of the Virgo cluster and beyond. Time is therefore ripe to develop the techniques required to extract the information available in such data. 

It should be emphasized that the use of star clusters as probes of  stellar populations  does come with its own set of caveats. The relative numbers of globular clusters and field stars vary strongly, both from galaxy to galaxy and within galaxies \citep{Harris1991,Harris2002,Brodie2006}. The  reasons for these variations  are poorly understood, but are likely due to a combination of dynamical evolutionary effects and variations in the formation efficiency of massive star clusters relative to field stars. While clusters may be used to identify bursts of star formation, e.g., due to major mergers, inferring the intensities of such bursts is therefore far less straight forward. 
In the context of chemical abundance analysis, it is also important to note that some elements are known to exhibit peculiar abundance patterns in globular clusters. These anomalies mostly affect the light elements  
He, C, N, O, Na, Mg and Al, and suggest that self-enrichment has taken place within globular clusters \citep{Gratton2012}. 
However, apart from these specific elements, the abundance patterns of globular clusters are generally similar to those seen in field stars, and at least in the Galactic disc both open clusters and field stars paint a consistent picture of age-metallicity relations and radial gradients \citep{Friel1995}.

Compared to  high-dispersion spectroscopy of individual stars, detailed abundance analysis from the integrated spectra of star clusters introduces  a number of additional challenges: First, the integrated optical light generally has significant contributions from stars across the Hertzsprung-Russell (H-R) diagram. In some cases this complication can be circumvented by choosing the wavelength range carefully. In particular, the near-infrared spectra of young ($\la100$ Myr) star clusters are dominated by red supergiant stars, and for such objects it may suffice to model the IR spectrum of an entire cluster as that of an individual, red supergiant \citep{Larsen2006b,Larsen2008a,Davies2010}. Second, although the  features in spectra of star clusters are much less broadened by internal velocity dispersions compared to, say, giant elliptical galaxies, the broadening can still lead to significant blending of spectral features. Measurements of the equivalent widths of individual lines therefore become more problematic than for single stars, requiring an increased reliance upon spectral synthesis. 
In recent years, there has been a general move  towards computing SSP models at increasing spectral resolution and wider wavelength coverage, and  grids of such models are available \citep{Delgado2005,Coelho2007}. However, for detailed abundance analysis it becomes impractical to rely on such pre-calculated grids, which would have to be tabulated for a large range of ages, overall metallicities, horizontal-branch morphologies, detailed chemical abundances, etc., and at a sufficiently fine spacing to allow constraints on the abundance patterns at the $\sim0.1$ dex level or better.

Our aim has been to develop a general method that is not so specialized that it is just  applicable  to high-dispersion, high S/N data of simple stellar populations with relatively modest internal velocity dispersions. In future applications, we may also apply it to  less ideal cases, such as data from spectrographs that deliver somewhat lower resolution (but are highly efficient, e.g., X-Shooter on the ESO VLT).
Conceptually, our approach has several aspects in common with that introduced by \citet{McWilliam2008} and further developed by \citet{Colucci2009,Colucci2011a,Colucci2012}. In both cases, the integrated light measurements are interpreted based on the detailed H-R diagrams of the clusters, combined with model atmosphere calculations and synthetic spectra of the kind that have traditionally been used for abundance analysis of individual stars.  However, we model larger sections of the integrated spectra, solving simultaneously for  (in principle) arbitrary combinations of individual element abundances. 
This allows us to use  information contained in many weak lines that may not be measurable individually, but still   constrain the abundance measurements.
A further motivation is that we may also, in later applications,  apply this type of analysis to medium-resolution spectra that will be more strongly affected by blending, especially for higher metallicities. This will clearly involve trade-offs in the level of detailed information that can be extracted from the data,  but we believe that the detailed spectral synthesis approach still has the potential to extract much more detailed information compared to, e.g., Lick/IDS index measurements. 

In this paper we introduce our approach and apply it to three of the globular clusters in the Fornax dwarf spheroidal galaxy. We start by briefly discussing previous work on the metallicities of the Fornax GCs (Sect.~\ref{sec:fornaxgc}). Our spectroscopic observations, their basic reduction, and the derivation of physical parameters for stars in the clusters are then discussed in Sect.~\ref{sec:uves_hst}, followed by a detailed description of the procedure by which we derive abundances from integrated-light spectroscopy (Sect.~\ref{sec:analysis}). In Sect.~\ref{sec:results} we present our results and carry out a number of tests to assess the various uncertainties.  In Sect.~\ref{sec:discussion} we discuss our measurements in the context of other abundance determinations for metal-poor stars in dwarf galaxies and the Milky Way, and our main conclusions are summarized in Sect.~\ref{sec:conclusions}.

\section{Globular clusters in the Fornax dwarf spheroidal as  a test case}
\label{sec:fornaxgc}

The globular clusters Fornax 3, Fornax 4 and Fornax 5 in the Fornax dwarf spheroidal galaxy \citep{Hodge1961} were chosen as an ideal first test sample for several reasons: 1) With core radii of $2\arcsec$--$3\arcsec$ and half-light radii of $5\arcsec$--$10\arcsec$ \citep{Mackey2003a}, they are sufficiently compact that integrated-light spectra can be obtained relatively easily. 2) At the same time, they are sufficiently nearby that resolved photometry from the Hubble Space Telescope (HST), reaching below the main sequence turn-off, is available \citep{Buonanno1998,Buonanno1999}.
Finally, although this was not known at the time our observations were planned, detailed abundance analysis has been carried out for \emph{individual} red giant branch (RGB) stars in one of the clusters \citep[Fornax 3;][]{Letarte2006},  providing a welcome comparison for  our integrated-light results.

Previous determinations of the metallicities of the Fornax GCs have been extensively reviewed and discussed by other authors \citep{Buonanno1998,Buonanno1999,Strader2003,Letarte2006,Mottini2008}. 
While most studies agree that Fornax 1, 2, 3 and 5 are very metal-poor with $\mathrm{[Fe/H]}\sim-2$, there are still substantial uncertainties. 
Past studies have relied on a variety of techniques, including  integrated colours, resolved photometry and low- or moderate dispersion integrated spectroscopy, most of which require intermediate calibration steps that are not always well established at such low metallicities. 
From measurements of Lick/IDS indices on Keck/LRIS spectra, \citet{Strader2003} derived somewhat higher metallicities of $\mathrm{[Fe/H]}\sim-1.8$ for Fornax 2, 3 and 5. The  UVES spectroscopy of \citet[][hereafter L06]{Letarte2006} instead yielded very low metallicities of $\mathrm{[Fe/H]}=-2.5$, $-2.1$ and $-2.4$ (with estimated $\pm0.1$ dex errors) for a few individual red giants in Fornax 1, 2 and 3, respectively. 

The case of \object{Fornax 4} is somewhat more controversial. Generally, this cluster has been thought to be  more metal-rich than the other four, with typical values of $\mathrm{[Fe/H]}\sim-1.4$ being derived from integrated photometry and/or low-dispersion spectroscopy \citep[see references in][]{Buonanno1999,Strader2003}. \citet{Strader2003} found $\mathrm{[Fe/H]}=-1.5$ for \object{Fornax 4}, and argued (given their relatively high metallicity estimates for the other clusters) that the difference between \object{Fornax 4} and the other GCs might be  smaller than suggested by previous studies.
From HST photometry,  \citet{Buonanno1999} also concluded that the metallicity of Fornax 4 is not significantly higher than that of the other clusters. In their case, this conclusion was based on their finding that the RGB slope was similar to that of the other Fornax GCs, suggesting a similar (low) metallicity of $\mathrm{[Fe/H]}\sim-2$. They did note that this was in disagreement with spectroscopic observations available at the time, and that the red horizontal branch of Fornax 4 would also be atypical of such a metal-poor cluster although this might be in part due to a younger age. 

Finally, we note that \object{Fornax 5} is known to host a  planetary nebula that was discovered in the same data used in this paper \citep{Larsen2008}. The PN is peculiar in having no detectable H Balmer emission lines, in spite of strong [\ion{O}{iii}] emission, and is one of only two known PNe in the Fornax dSph.

\section{Observations and data analysis}
\label{sec:uves_hst}

\subsection{UVES spectroscopy}

\begin{figure}
\includegraphics[width=\columnwidth]{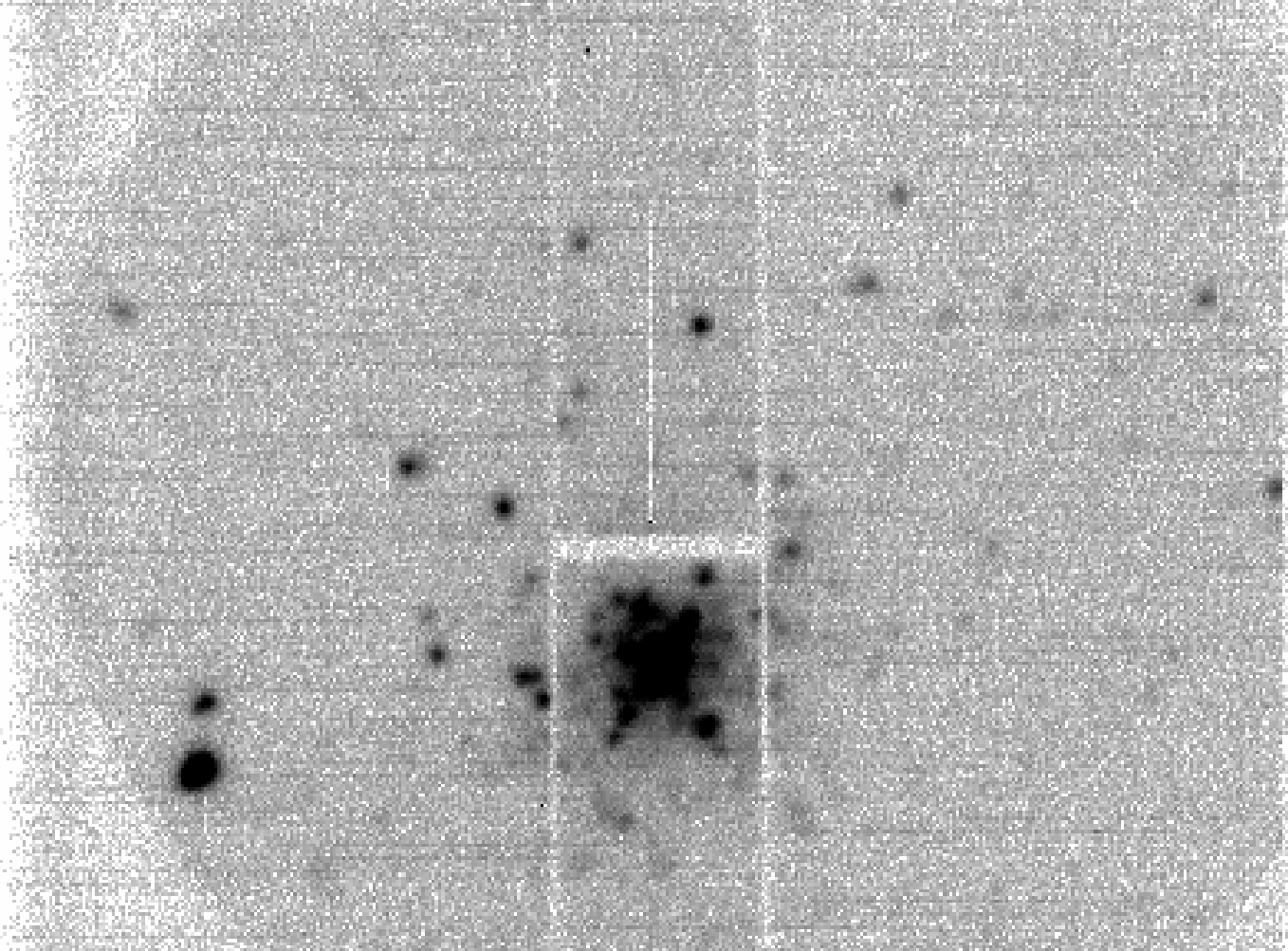}
\caption{\label{fig:F4sv}Image of Fornax 4 from the UVES slit-viewing camera at the start of an East--West scan. The 10$\arcsec$ long UVES slit  is  visible at its initial position, one half-light radius ($5\farcs5$) from the centre of the cluster.}
\end{figure}

We obtained integrated-light spectra of Fornax 3, 4 and 5 with the UVES spectrograph \citep{Dekker2000} on the ESO Very Large Telescope on 19 Nov and 20 Nov, 2006. The observations were made with the UVES red arm, using cross-disperser \#3 and a slit width of $1\arcsec$. This set-up yielded a resolving power $R\sim40\,000$ and a
spectral range of 4200~\AA\ -- 6200~\AA, with a gap from 5170~\AA\ -- 5230~\AA\ between the two CCD detectors. 
The seeing varied between $0\farcs5$ and $1\farcs5$ during the two nights, but the actual seeing was of little importance for our purpose. Weather conditions were mostly clear, with some clouds near the horizon that did not significantly affect our observations.

To sample the integrated light, we scanned the UVES slit across each cluster in both the North--South and East--West directions. Each scan had an exposure time of 2400 s, and during this time the slit was moved across the half-light diameter of the cluster. We used the structural parameters derived from HST imaging by \citet{Mackey2003a}, according to which  the clusters have half-light radii of $r_\mathrm{eff} = 8\farcs2$ (\object{Fornax~3}), $r_\mathrm{eff} = 5\farcs5$ (\object{Fornax~4}) and $r_\mathrm{eff} = 9\farcs6$ (\object{Fornax~5}). Two 2400 s exposures were obtained in each direction,  leading to a total (science) exposure time of 9600 s for each cluster.
Figure~\ref{fig:F4sv} shows the acquisition image for one of the Fornax~4 spectra. Even this cluster, located near the centre of the Fornax dSph,  clearly stands out above the background.
However, because of the limited length of the UVES slit (10$\arcsec$) separate sky exposures were required  to properly subtract the sky background from the science spectra. These were obtained by scanning the slit parallel to the science exposures, but at offsets of $2\farcm5$ from each cluster. Each sky scan had an exposure time of 1200 s, and two such scans bracketing each science scan were co-added and used as a sky reference for the science exposures. We verified that these sky exposures were made at locations where no bright, individual stars were evident.

\begin{figure}
\includegraphics[width=\columnwidth]{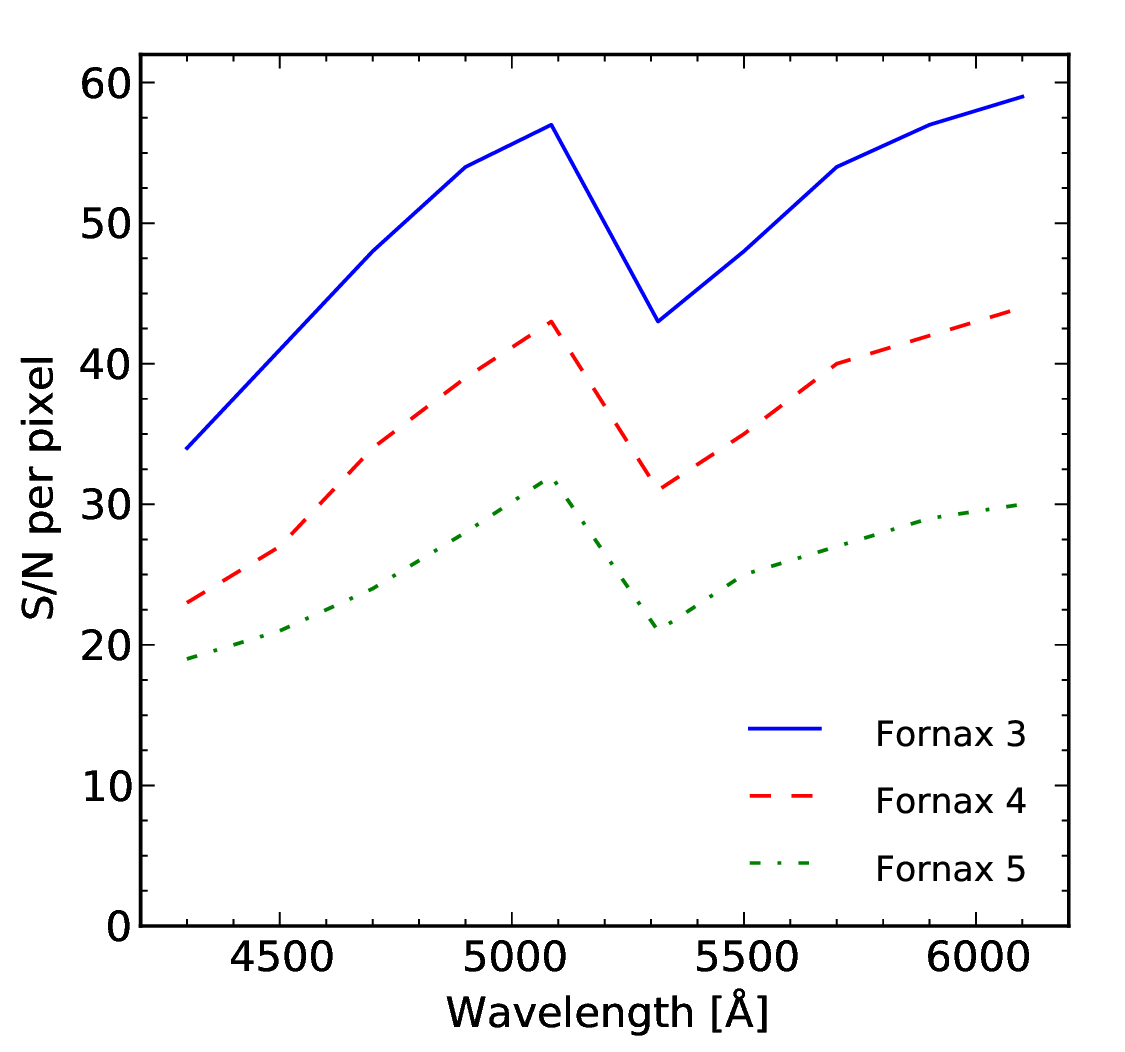}
\caption{\label{fig:s2n}Signal-to-noise per (re-binned) pixel versus wavelength. Below 5200~\AA, one pixel corresponds to 0.025~\AA; above 5200~\AA\ a pixel spans 0.030~\AA .}
\end{figure}

Initial reduction of the spectra was carried out with version 3.3.1 of the standard UVES pipeline provided by ESO, running within the \texttt{Gasgano} environment. The pipeline performs bias- and inter-order background subtraction and flat-fields the spectra. It then carries out wavelength calibration (using calibration lamp spectra) and rebins the echelle orders to a common spatial scale. Finally, the echelle orders are merged and a single two-dimensional, wavelength calibrated spectrum is output for each of the two CCD detectors. Using standard tools in the \texttt{IRAF}\footnote{IRAF is distributed
by the National Optical Astronomical Observatories, which are operated by the Association of Universities for Research in Astronomy, Inc.~under contract with the National Science Foundation} 
package, we then co-added the two sky exposures corresponding to each science exposure, subtracted the sky from the science data, and then co-added the four individual scans for each cluster. Prior to the pipeline reduction, minor detector blemishes  were repaired with the \texttt{fixpix} task in \texttt{IRAF}, but the red CCD has a group of hot columns that remain visible as an artifact that repeats every $\sim50$~\AA\ in the final spectra. The 2-D spectra were finally collapsed to one-dimensional spectra and bad regions were masked out (assigned zero weight)  for further analysis. Fig.~\ref{fig:s2n} shows the signal-to-noise ratio (averaged over 200~\AA\ intervals) as a function of wavelength for each cluster.  The discontinuity at 5200~\AA\ corresponds to the transition from the ``blue'' (EEV) to the ``red'' (MIT-LL) CCD detector.

\begin{figure}
\centering
\includegraphics[width=\columnwidth]{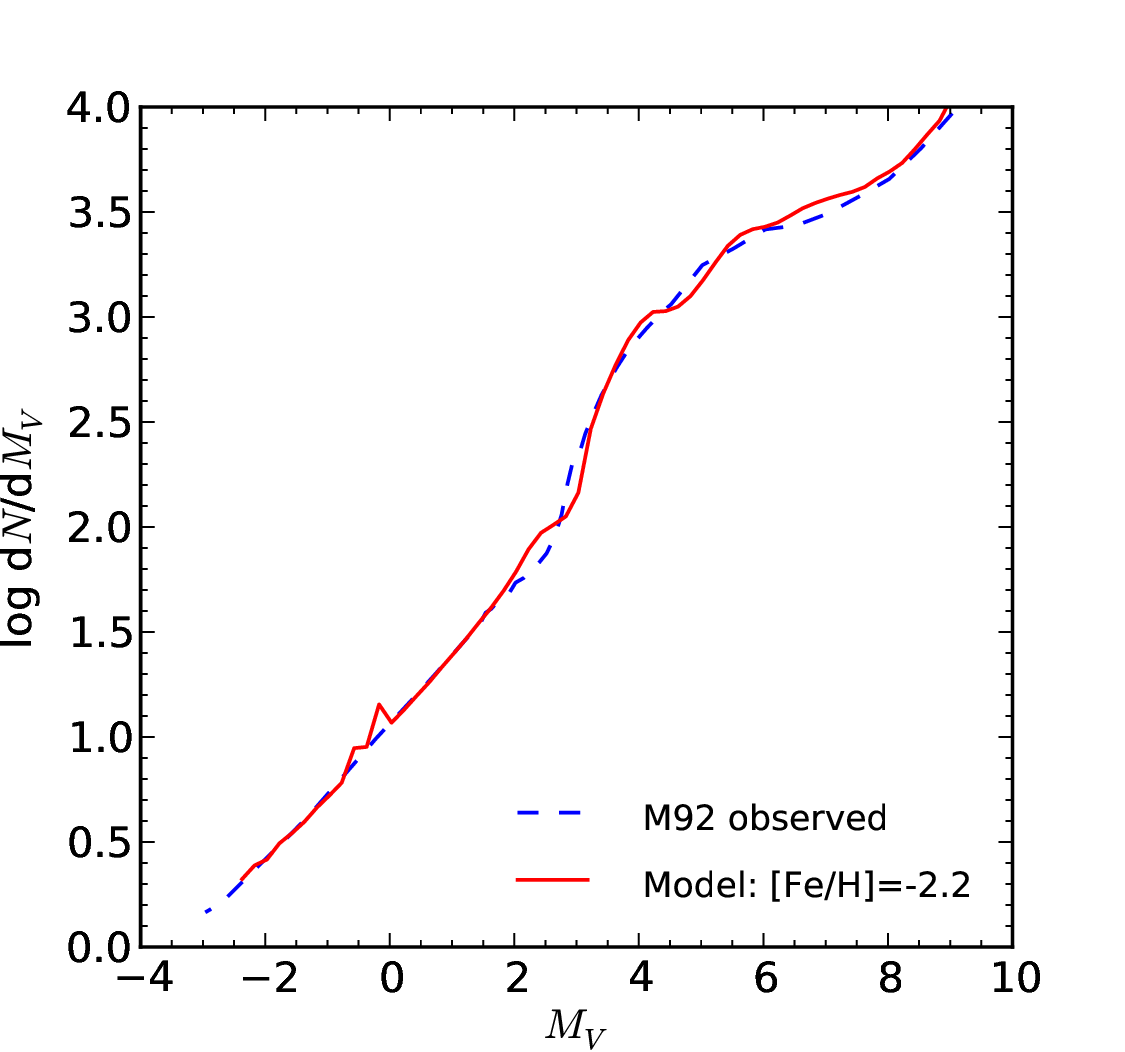}
\caption{\label{fig:lfs}An empirical luminosity function for the globular cluster M92, compared with a model from \citet{Dotter2007}.
}
\end{figure}

\begin{figure*}
\centering
\includegraphics[width=60mm]{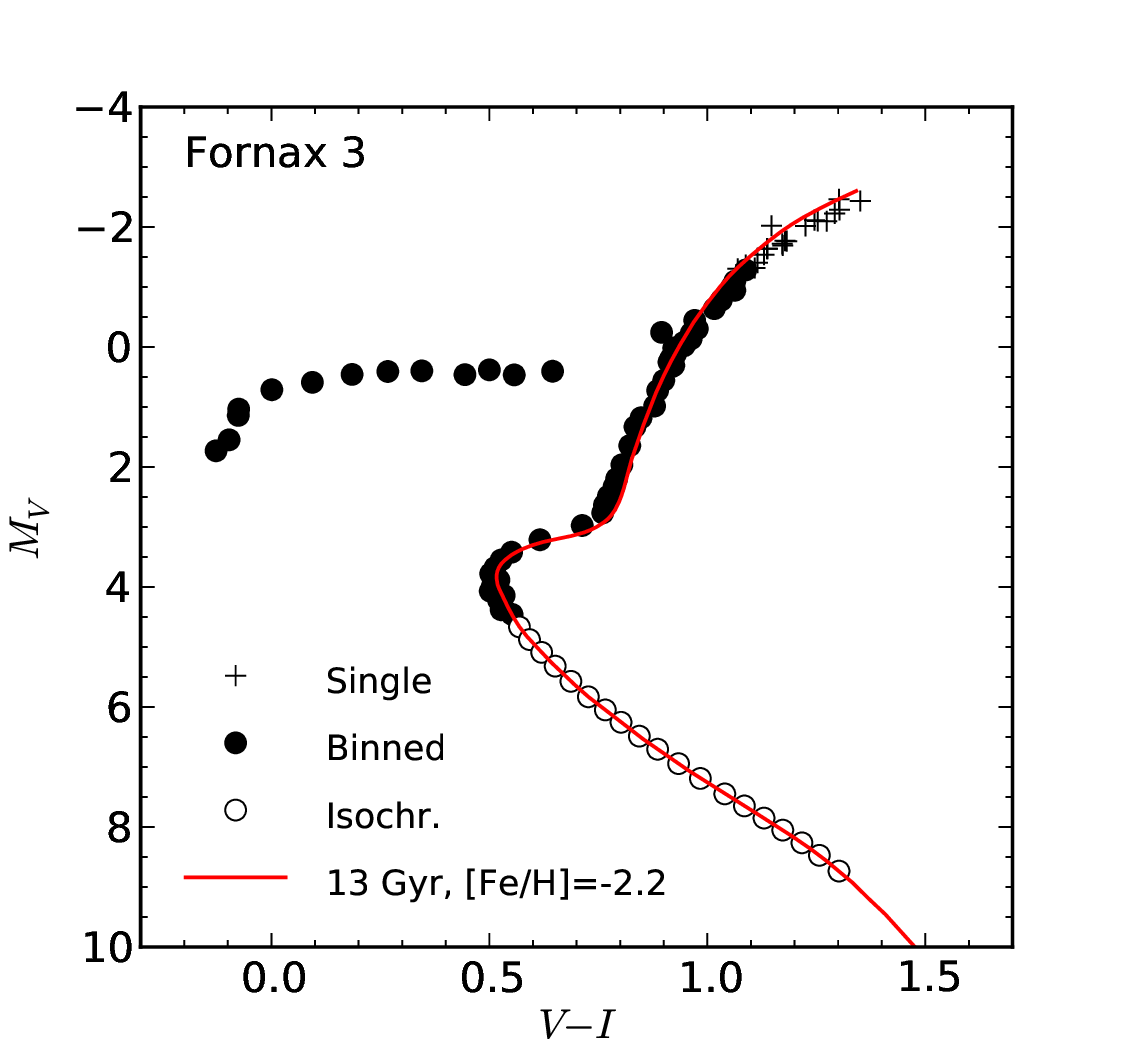}
\includegraphics[width=60mm]{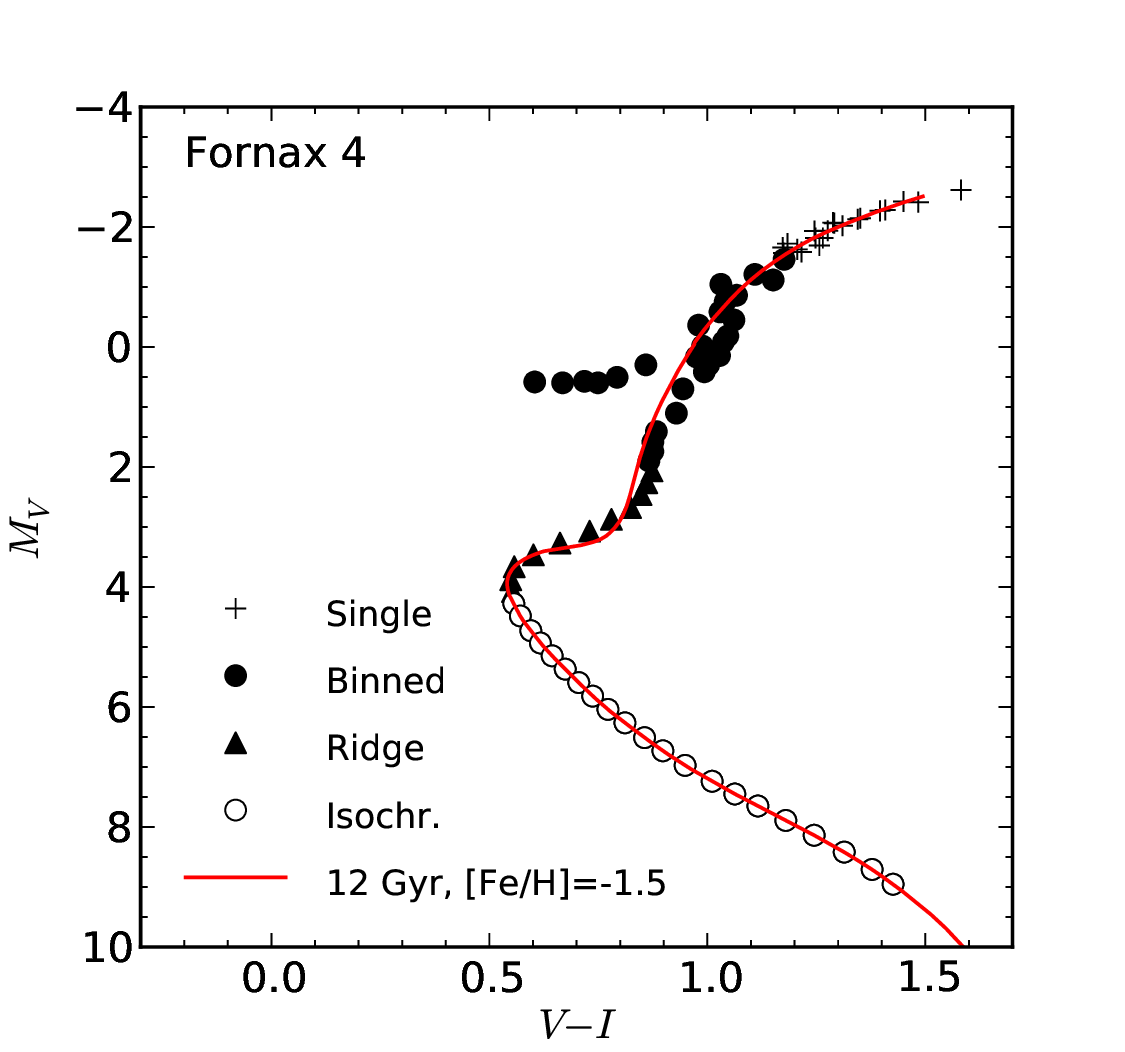}
\includegraphics[width=60mm]{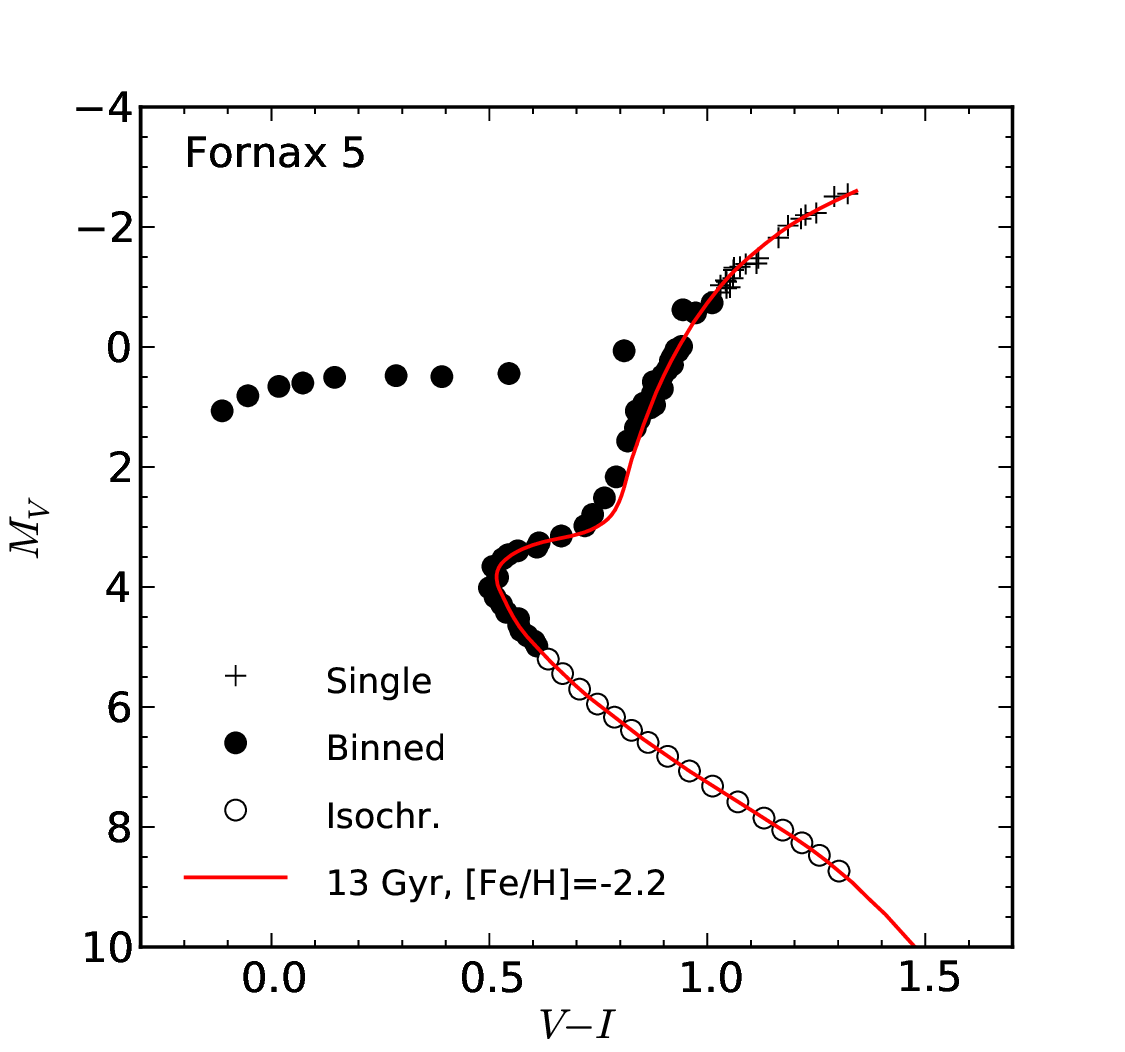}
\caption{\label{fig:cmds}Binned colour-magnitude diagrams used in the generation of synthetic integrated-light spectra. The plus markers denote individual stars, filled circles are bins of multiple  stars, and open squares are model extrapolations from the \citet{Dotter2007} isochrones. For Fornax 4, the \citet{Buonanno1999} ridge line was used near the main sequence turn-off and for the  subgiants, shown with filled triangles. Theoretical isochrones used to model the luminosity functions \citep{Dotter2007} are shown as red curves.
 }
\end{figure*}

\subsection{Colour-magnitude diagrams and luminosity functions}
\label{sec:cmds}

To derive stellar parameters for \object{Fornax 3} and \object{Fornax 5} we used the HST-based photometry of \citet{Buonanno1998}, kindly provided to us by C.\ E.\ Corsi (priv.\ comm.).  For \object{Fornax 4} we used the photometry from \citet{Buonanno1999}, available through the CDS archive\footnote{http://cdsarc.u-strasbg.fr/}. We transformed the photometry in the instrumental (WFPC2) system to standard Johnson-Cousins $V$ and $I$ magnitudes using the  relations in \citet{Holtzman1995}. The  colour-magnitude diagrams (CMDs) of \object{Fornax 3} and \object{Fornax 5} reach a limit of $M_V\approx+5$, just below the main sequence turn-off. Still, stars below this limit contribute about 25\% of the $V$-band luminosity (for the luminosity functions shown in Fig.~\ref{fig:lfs}; see discussion below), a non-negligible fraction. 
The data are somewhat shallower for  \object{Fornax 4} (reaching to $M_V\approx+4$) which is located in the more crowded central regions of the galaxy and was imaged on the WFPC2 WF3 chip (with a pixel scale of $0\farcs10$), rather than the PC detector ($0\farcs045$). We  extrapolated below the observational limits using theoretical isochrones that have been extensively tested against deep HST observations of Galactic globular clusters \citep{Dotter2007}.

To properly model the integrated light, we need to know not only the \emph{shape} of the CMDs, but also the relative numbers of stars populating each region of the CMDs. Here, it was less straight-forward to use the HST photometry since it is based on a combination of two datasets (a ``deep'' and a ``shallow'' exposure) for each cluster that cover different radial ranges. In addition, no detailed information is available about the completeness of the photometry. 
For the luminosity functions (LFs) we therefore relied more heavily on the theoretical models of \citet{Dotter2007}.
Fig.~\ref{fig:lfs} shows a comparison of empirical and model LFs for the metal-poor Galactic globular cluster M92. The empirical LF was obtained by combining the data of \citet{Lee2003} and \citet{Paust2007}, while the model LF  assumes a \citet{Salpeter1955} stellar mass function (i.e.\ a uniform power-law $dN/dM \propto M^{-2.35}$). Overall, the agreement is very good within the range of absolute magnitudes for which empirical data are available (i.e., down to $M_V=+9$). 
For stars brighter than the main sequence turn-off, the LF shape is mostly determined by stellar evolution and does not depend significantly on the assumed stellar mass function.  At masses below $\sim0.5 M_\odot$, corresponding to $M_V \ga +8$,  the mass function in globular clusters (and hence the LF) is in general significantly shallower than the Salpeter law \citep{Bastian2010,Paresce2000,DeMarchi2010}, so an LF based on a pure Salpeter law will overestimate the contribution from stars below this limit.  The detailed shape of the low-mass end of the MF might be estimated if the degree of dynamical evolution experienced by each cluster were known \citep[e.g.][]{Kruijssen2009a}, but this is beyond the scope of this work. Given the uncertainty about the  MF  at the low-mass end, we simply adopted the \citet{Dotter2007} LF down to a magnitude limit of $M_V\sim+9$ and ignored fainter stars. Stars below this limit  do contribute to the mass (and hence affect the $M/L$ ratio), but their contribution to the integrated light is small, less than 5\% (in the $V$-band) if the Salpeter slope extends down to the lowest masses, and probably less than that for any real cluster. Nevertheless, the behaviour of the low-mass end of the LF is a significant contributor to the overall uncertainties on our abundance determinations for some elements (Sect.~\ref{sec:uncertainties}).

It is not desirable to rely fully on theoretical LFs, particularly for the horizontal branch (HB). We therefore adopted a hybrid approach in which we based our modelling of the integrated light purely on the observed CMDs for the brighter stars (including the HB), combined with the theoretical LFs for fainter stars. For Fornax 3 and 5 we used LFs computed for [Fe/H]=$-$2.2, an age of 13 Gyr and [$\alpha$/Fe]=$+0.2$, while we assumed [Fe/H]=$-$1.5, an age of 12 Gyr and [$\alpha$/Fe]=0.0 for Fornax 4. 
It is important to note that the switch from  empirical information to theoretical modelling occurs at a significantly brighter threshold for the LFs than for the overall CMD shape. Specifically, we used empirical LF information for red giant branch (RGB) stars brighter than $M_V=0$ and for the HB, while the Dotter et al. LFs were used elsewhere. These limits are, of course, specific to the dataset used here, and would be different in other applications of our technique.
Note that the stars used in the CMD modelling were not selected to fall strictly within the area covered by the slit scans. The photometry extends out to  larger radii ($\sim15\arcsec$) than those included in the scans, but does not include stars closer than $\sim2\arcsec$ to the centre. Our modelling of the integrated light is therefore subject to stochastic (and possibly systematic) variations in the number of stars of different types within the slit scan areas. The potential impact of these effects is discussed in detail in Sect.~\ref{sec:uncertainties}.

In some globular clusters, blue stragglers (BSs)  contribute significantly to the integrated light.  The CMDs of the three GCs studied here do not show any significant population of blue stragglers, and they are not included in the modelling. Inclusion of BSs  would, in principle, be straight forward for clusters where the location of the BSs in the H-R diagram and their relative numbers  are known. However, for clusters with significant BS populations a potential complication is that BSs may exhibit peculiar abundance patterns \citep{Lovisi2012}.

The foreground extinction towards the \object{Fornax dSph} is small but non-negligible. For \object{Fornax 3} and \object{Fornax 5} we assumed $E(V\!-\!I)=0.05$ mag and $E(V\!-\!I)=0.08$ mag and a distance modulus $(m-M)_0=20.68$ mag \citep{Buonanno1998}. For \object{Fornax 4} there is a larger range of $E(V\!-\!I)$ values quoted in the literature. \citet{Buonanno1999} find a relatively large foreground reddening of $E(V\!-\!I)=0.15$ mag, while \citet{Greco2007} find $E(B\!-\!V)=0.08\pm0.03$ from the blue edge of the RR Lyrae strip, corresponding to $E(V\!-\!I) = 0.093$. For the larger reddening value of $E(V\!-\!I)=0.15$, the RGB ridge line of \object{Fornax 4} becomes very similar to that of the other Fornax GCs.
The smaller $E(V\!-\!I)$ value of \citet{Greco2007} leads to an intrinsically redder RGB, more consistent with a higher metallicity.
In this paper we will adopt $E(V\!-\!I)=0.09$ for \object{Fornax 4}, along with the same distance modulus as for the other two clusters. Varying the distance modulus within the range found by various authors ($\sim0.1$ mag) has a negligible effect on our results. We will discuss the uncertainties related to the extinction correction in more detail below.

\onltab{1}{
\begin{longtable}{rrcccl}
\caption{\label{tab:cmdf3}Colour-magnitude diagram for Fornax 3}\\\hline\hline
$M_V$ & $V\!-\!I$ & Weight & & $f(L_V)$ & \\ \hline 
\endfirsthead
\caption{continued.}\\ \hline \hline 
$M_V$ & $V\!-\!I$ & Weight & & $f(L_V)$ & \\ \hline 
\endhead 
\hline 
\endfoot 
$ 1.035$ & $-0.075$ & 6 & E & 0.002 & EHB\\
$ 1.141$ & $-0.076$ & 6 & E & 0.005 & EHB\\
$ 1.547$ & $-0.097$ & 6 & E & 0.006 & EHB\\
$ 1.730$ & $-0.127$ & 6 & E & 0.007 & EHB\\
$ 0.712$ & $ 0.001$ & 17 & E & 0.016 & HB\\
$ 0.589$ & $ 0.094$ & 17 & E & 0.027 & HB\\
$ 0.457$ & $ 0.185$ & 17 & E & 0.038 & HB\\
$ 0.408$ & $ 0.267$ & 17 & E & 0.050 & HB\\
$ 0.398$ & $ 0.345$ & 17 & E & 0.062 & HB\\
$ 0.461$ & $ 0.444$ & 17 & E & 0.074 & HB\\
$ 0.381$ & $ 0.500$ & 17 & E & 0.086 & HB\\
$ 0.465$ & $ 0.557$ & 17 & E & 0.097 & HB\\
$ 0.404$ & $ 0.645$ & 17 & E & 0.109 & HB\\
$-0.244$ & $ 0.895$ & 17 & E & 0.131 & HB\\
$-2.462$ & $ 1.302$ & 1 & E & 0.141 & RGB/MS\\
$-2.434$ & $ 1.351$ & 1 & E & 0.151 & RGB/MS\\
$-2.286$ & $ 1.303$ & 1 & E & 0.159 & RGB/MS\\
$-2.224$ & $ 1.292$ & 1 & E & 0.167 & RGB/MS\\
$-2.114$ & $ 1.246$ & 1 & E & 0.175 & RGB/MS\\
$-2.100$ & $ 1.253$ & 1 & E & 0.182 & RGB/MS\\
$-2.095$ & $ 1.274$ & 1 & E & 0.189 & RGB/MS\\
$-2.022$ & $ 1.147$ & 1 & E & 0.196 & RGB/MS\\
$-2.016$ & $ 1.225$ & 1 & E & 0.202 & RGB/MS\\
$-1.770$ & $ 1.178$ & 1 & E & 0.207 & RGB/MS\\
$-1.764$ & $ 1.182$ & 1 & E & 0.213 & RGB/MS\\
$-1.719$ & $ 1.172$ & 1 & E & 0.218 & RGB/MS\\
$-1.692$ & $ 1.173$ & 1 & E & 0.223 & RGB/MS\\
$-1.639$ & $ 1.137$ & 1 & E & 0.227 & RGB/MS\\
$-1.637$ & $ 1.138$ & 1 & E & 0.232 & RGB/MS\\
$-1.537$ & $ 1.130$ & 1 & E & 0.236 & RGB/MS\\
$-1.403$ & $ 1.115$ & 1 & E & 0.240 & RGB/MS\\
$-1.366$ & $ 1.088$ & 1 & E & 0.243 & RGB/MS\\
$-1.317$ & $ 1.109$ & 1 & E & 0.247 & RGB/MS\\
$-1.304$ & $ 1.070$ & 1 & E & 0.250 & RGB/MS\\
$-1.284$ & $ 1.088$ & 5 & E & 0.267 & RGB/MS\\
$-1.096$ & $ 1.063$ & 5 & E & 0.281 & RGB/MS\\
$-1.000$ & $ 1.056$ & 5 & E & 0.294 & RGB/MS\\
$-0.945$ & $ 1.063$ & 5 & E & 0.306 & RGB/MS\\
$-0.780$ & $ 1.032$ & 5 & E & 0.317 & RGB/MS\\
$-0.639$ & $ 1.016$ & 5 & E & 0.326 & RGB/MS\\
$-0.444$ & $ 0.971$ & 5 & E & 0.334 & RGB/MS\\
$-0.304$ & $ 0.977$ & 5 & E & 0.341 & RGB/MS\\
$-0.227$ & $ 0.963$ & 5 & E & 0.347 & RGB/MS\\
$-0.179$ & $ 0.962$ & 5 & E & 0.353 & RGB/MS\\
$-0.138$ & $ 0.963$ & 5 & E & 0.359 & RGB/MS\\
$-0.097$ & $ 0.954$ & 5 & E & 0.365 & RGB/MS\\
$-0.077$ & $ 0.946$ & 5 & E & 0.370 & RGB/MS\\
$-0.021$ & $ 0.947$ & 5 & E & 0.376 & RGB/MS\\
$ 0.016$ & $ 0.923$ & 5 & E & 0.381 & RGB/MS\\
$ 0.118$ & $ 0.927$ & 5 & E & 0.385 & RGB/MS\\
$ 0.193$ & $ 0.916$ & 5 & E & 0.390 & RGB/MS\\
$ 0.252$ & $ 0.912$ & 5 & E & 0.394 & RGB/MS\\
$ 0.307$ & $ 0.923$ & 5 & E & 0.398 & RGB/MS\\
$ 0.327$ & $ 0.919$ & 5 & E & 0.401 & RGB/MS\\
$ 0.557$ & $ 0.900$ & 3.55e+01 & S & 0.423 & RGB/MS\\
$ 0.726$ & $ 0.886$ & 2.63e+01 & S & 0.437 & RGB/MS\\
$ 0.984$ & $ 0.879$ & 3.03e+01 & S & 0.450 & RGB/MS\\
$ 1.179$ & $ 0.848$ & 3.69e+01 & S & 0.463 & RGB/MS\\
$ 1.330$ & $ 0.834$ & 3.46e+01 & S & 0.473 & RGB/MS\\
$ 1.644$ & $ 0.822$ & 1.00e+02 & S & 0.496 & RGB/MS\\
$ 1.963$ & $ 0.804$ & 1.17e+02 & S & 0.515 & RGB/MS\\
$ 2.195$ & $ 0.792$ & 8.26e+01 & S & 0.527 & RGB/MS\\
$ 2.327$ & $ 0.786$ & 8.17e+01 & S & 0.537 & RGB/MS\\
$ 2.486$ & $ 0.773$ & 8.97e+01 & S & 0.546 & RGB/MS\\
$ 2.632$ & $ 0.764$ & 9.89e+01 & S & 0.555 & RGB/MS\\
$ 2.765$ & $ 0.760$ & 7.52e+01 & S & 0.561 & RGB/MS\\
$ 2.975$ & $ 0.713$ & 2.62e+02 & S & 0.578 & RGB/MS\\
$ 3.213$ & $ 0.616$ & 4.22e+02 & S & 0.601 & RGB/MS\\
$ 3.420$ & $ 0.551$ & 4.90e+02 & S & 0.623 & RGB/MS\\
$ 3.550$ & $ 0.527$ & 4.07e+02 & S & 0.638 & RGB/MS\\
$ 3.677$ & $ 0.513$ & 4.81e+02 & S & 0.655 & RGB/MS\\
$ 3.779$ & $ 0.503$ & 4.65e+02 & S & 0.670 & RGB/MS\\
$ 3.883$ & $ 0.522$ & 5.92e+02 & S & 0.687 & RGB/MS\\
$ 3.982$ & $ 0.507$ & 5.86e+02 & S & 0.702 & RGB/MS\\
$ 4.069$ & $ 0.502$ & 4.68e+02 & S & 0.714 & RGB/MS\\
$ 4.142$ & $ 0.534$ & 4.82e+02 & S & 0.725 & RGB/MS\\
$ 4.220$ & $ 0.521$ & 5.32e+02 & S & 0.736 & RGB/MS\\
$ 4.294$ & $ 0.528$ & 5.63e+02 & S & 0.747 & RGB/MS\\
$ 4.379$ & $ 0.527$ & 5.65e+02 & S & 0.757 & RGB/MS\\
$ 4.456$ & $ 0.552$ & 5.14e+02 & S & 0.766 & RGB/MS\\
$ 4.665$ & $ 0.569$ & 1.57e+03 & S & 0.788 & ISO\\
$ 4.876$ & $ 0.592$ & 1.82e+03 & S & 0.809 & ISO\\
$ 5.090$ & $ 0.620$ & 2.24e+03 & S & 0.830 & ISO\\
$ 5.319$ & $ 0.651$ & 2.98e+03 & S & 0.853 & ISO\\
$ 5.572$ & $ 0.687$ & 3.95e+03 & S & 0.877 & ISO\\
$ 5.830$ & $ 0.727$ & 4.44e+03 & S & 0.899 & ISO\\
$ 6.049$ & $ 0.766$ & 3.87e+03 & S & 0.914 & ISO\\
$ 6.255$ & $ 0.802$ & 3.84e+03 & S & 0.926 & ISO\\
$ 6.486$ & $ 0.844$ & 4.73e+03 & S & 0.939 & ISO\\
$ 6.704$ & $ 0.886$ & 4.80e+03 & S & 0.949 & ISO\\
$ 6.944$ & $ 0.934$ & 5.65e+03 & S & 0.959 & ISO\\
$ 7.191$ & $ 0.984$ & 6.12e+03 & S & 0.967 & ISO\\
$ 7.448$ & $ 1.040$ & 6.69e+03 & S & 0.974 & ISO\\
$ 7.649$ & $ 1.085$ & 5.54e+03 & S & 0.979 & ISO\\
$ 7.851$ & $ 1.130$ & 6.13e+03 & S & 0.984 & ISO\\
$ 8.052$ & $ 1.173$ & 6.59e+03 & S & 0.988 & ISO\\
$ 8.264$ & $ 1.217$ & 7.75e+03 & S & 0.992 & ISO\\
$ 8.472$ & $ 1.257$ & 8.87e+03 & S & 0.995 & ISO\\
$ 8.735$ & $ 1.302$ & 1.39e+04 & S & 1.000 & ISO\\
\hline
\end{longtable}
}

\onltab{2}{
\begin{longtable}{rrcccl}
\caption{\label{tab:cmdf4}Colour-magnitude diagram for Fornax 4}\\\hline\hline
$M_V$ & $V\!-\!I$ & Weight & & $f(L_V)$ & \\ \hline 
\endfirsthead
\caption{continued.}\\ \hline \hline 
$M_V$ & $V\!-\!I$ & Weight & & $f(L_V)$ & \\ \hline 
\endhead 
\hline 
\endfoot 
$ 0.583$ & $ 0.604$ & 14 & E & 0.009 & HB\\
$ 0.597$ & $ 0.668$ & 14 & E & 0.019 & HB\\
$ 0.572$ & $ 0.718$ & 14 & E & 0.028 & HB\\
$ 0.597$ & $ 0.749$ & 14 & E & 0.038 & HB\\
$ 0.505$ & $ 0.793$ & 14 & E & 0.048 & HB\\
$ 0.298$ & $ 0.859$ & 4 & E & 0.051 & HB\\
$-2.617$ & $ 1.582$ & 1 & E & 0.064 & RGB/MS\\
$-2.426$ & $ 1.450$ & 1 & E & 0.075 & RGB/MS\\
$-2.414$ & $ 1.484$ & 1 & E & 0.086 & RGB/MS\\
$-2.283$ & $ 1.408$ & 1 & E & 0.095 & RGB/MS\\
$-2.269$ & $ 1.396$ & 1 & E & 0.104 & RGB/MS\\
$-2.147$ & $ 1.351$ & 1 & E & 0.113 & RGB/MS\\
$-2.129$ & $ 1.345$ & 1 & E & 0.121 & RGB/MS\\
$-2.073$ & $ 1.288$ & 1 & E & 0.129 & RGB/MS\\
$-2.064$ & $ 1.291$ & 1 & E & 0.136 & RGB/MS\\
$-2.021$ & $ 1.310$ & 1 & E & 0.144 & RGB/MS\\
$-1.937$ & $ 1.276$ & 1 & E & 0.151 & RGB/MS\\
$-1.934$ & $ 1.246$ & 1 & E & 0.158 & RGB/MS\\
$-1.816$ & $ 1.265$ & 1 & E & 0.164 & RGB/MS\\
$-1.817$ & $ 1.248$ & 1 & E & 0.170 & RGB/MS\\
$-1.723$ & $ 1.184$ & 1 & E & 0.175 & RGB/MS\\
$-1.693$ & $ 1.257$ & 1 & E & 0.181 & RGB/MS\\
$-1.657$ & $ 1.173$ & 1 & E & 0.186 & RGB/MS\\
$-1.627$ & $ 1.206$ & 1 & E & 0.191 & RGB/MS\\
$-1.584$ & $ 1.216$ & 1 & E & 0.196 & RGB/MS\\
$-1.568$ & $ 1.175$ & 1 & E & 0.201 & RGB/MS\\
$-1.461$ & $ 1.176$ & 5 & E & 0.223 & RGB/MS\\
$-1.210$ & $ 1.109$ & 5 & E & 0.241 & RGB/MS\\
$-1.117$ & $ 1.151$ & 5 & E & 0.257 & RGB/MS\\
$-1.045$ & $ 1.031$ & 5 & E & 0.272 & RGB/MS\\
$-0.862$ & $ 1.067$ & 5 & E & 0.285 & RGB/MS\\
$-0.758$ & $ 1.041$ & 5 & E & 0.297 & RGB/MS\\
$-0.584$ & $ 1.029$ & 5 & E & 0.307 & RGB/MS\\
$-0.450$ & $ 1.061$ & 5 & E & 0.315 & RGB/MS\\
$-0.364$ & $ 0.980$ & 5 & E & 0.323 & RGB/MS\\
$-0.184$ & $ 1.047$ & 5 & E & 0.330 & RGB/MS\\
$-0.086$ & $ 1.037$ & 5 & E & 0.336 & RGB/MS\\
$-0.015$ & $ 0.989$ & 5 & E & 0.342 & RGB/MS\\
$ 0.143$ & $ 1.028$ & 5 & E & 0.347 & RGB/MS\\
$ 0.166$ & $ 0.975$ & 5 & E & 0.352 & RGB/MS\\
$ 0.266$ & $ 0.993$ & 5 & E & 0.357 & RGB/MS\\
$ 0.298$ & $ 1.003$ & 5 & E & 0.361 & RGB/MS\\
$ 0.413$ & $ 0.993$ & 2.15e+01 & S & 0.378 & RGB/MS\\
$ 0.698$ & $ 0.944$ & 4.67e+01 & S & 0.406 & RGB/MS\\
$ 1.103$ & $ 0.929$ & 6.52e+01 & S & 0.434 & RGB/MS\\
$ 1.409$ & $ 0.883$ & 3.79e+01 & S & 0.446 & RGB/MS\\
$ 1.584$ & $ 0.875$ & 4.49e+01 & S & 0.458 & RGB/MS\\
$ 1.740$ & $ 0.875$ & 4.65e+01 & S & 0.469 & RGB/MS\\
$ 1.898$ & $ 0.866$ & 5.00e+01 & S & 0.479 & RGB/MS\\
$ 2.066$ & $ 0.873$ & 7.30e+01 & S & 0.491 & RIDGE\\
$ 2.263$ & $ 0.861$ & 8.87e+01 & S & 0.504 & RIDGE\\
$ 2.460$ & $ 0.848$ & 1.13e+02 & S & 0.517 & RIDGE\\
$ 2.677$ & $ 0.824$ & 1.30e+02 & S & 0.530 & RIDGE\\
$ 2.874$ & $ 0.780$ & 1.41e+02 & S & 0.542 & RIDGE\\
$ 3.070$ & $ 0.730$ & 1.64e+02 & S & 0.553 & RIDGE\\
$ 3.268$ & $ 0.662$ & 2.68e+02 & S & 0.568 & RIDGE\\
$ 3.465$ & $ 0.601$ & 5.59e+02 & S & 0.595 & RIDGE\\
$ 3.662$ & $ 0.557$ & 6.84e+02 & S & 0.622 & RIDGE\\
$ 3.879$ & $ 0.549$ & 8.89e+02 & S & 0.651 & RIDGE\\
$ 4.076$ & $ 0.553$ & 1.09e+03 & S & 0.680 & RIDGE\\
$ 4.277$ & $ 0.556$ & 1.33e+03 & S & 0.710 & ISO\\
$ 4.482$ & $ 0.571$ & 1.62e+03 & S & 0.740 & ISO\\
$ 4.728$ & $ 0.595$ & 2.26e+03 & S & 0.774 & ISO\\
$ 4.933$ & $ 0.617$ & 2.14e+03 & S & 0.800 & ISO\\
$ 5.147$ & $ 0.644$ & 2.54e+03 & S & 0.825 & ISO\\
$ 5.369$ & $ 0.674$ & 2.98e+03 & S & 0.850 & ISO\\
$ 5.593$ & $ 0.705$ & 3.36e+03 & S & 0.872 & ISO\\
$ 5.819$ & $ 0.737$ & 3.72e+03 & S & 0.893 & ISO\\
$ 6.040$ & $ 0.772$ & 3.93e+03 & S & 0.910 & ISO\\
$ 6.264$ & $ 0.811$ & 4.22e+03 & S & 0.925 & ISO\\
$ 6.511$ & $ 0.856$ & 4.94e+03 & S & 0.939 & ISO\\
$ 6.730$ & $ 0.898$ & 4.54e+03 & S & 0.950 & ISO\\
$ 6.973$ & $ 0.949$ & 5.17e+03 & S & 0.960 & ISO\\
$ 7.238$ & $ 1.011$ & 5.82e+03 & S & 0.968 & ISO\\
$ 7.450$ & $ 1.063$ & 4.79e+03 & S & 0.974 & ISO\\
$ 7.650$ & $ 1.116$ & 4.66e+03 & S & 0.979 & ISO\\
$ 7.891$ & $ 1.180$ & 6.02e+03 & S & 0.983 & ISO\\
$ 8.138$ & $ 1.245$ & 6.89e+03 & S & 0.988 & ISO\\
$ 8.418$ & $ 1.314$ & 8.87e+03 & S & 0.992 & ISO\\
$ 8.708$ & $ 1.378$ & 1.12e+04 & S & 0.997 & ISO\\
$ 8.954$ & $ 1.426$ & 1.15e+04 & S & 1.000 & ISO\\
\hline
\end{longtable}
}

\onltab{3}{
\begin{longtable}{rrcccl}
\caption{\label{tab:cmdf5}Colour-magnitude diagram for Fornax 5}\\\hline\hline
$M_V$ & $V\!-\!I$ & Weight & & $f(L_V)$ & \\ \hline 
\endfirsthead
\caption{continued.}\\ \hline \hline 
$M_V$ & $V\!-\!I$ & Weight & & $f(L_V)$ & \\ \hline 
\endhead 
\hline 
\endfoot 
$ 1.064$ & $-0.113$ & 14 & E & 0.011 & HB\\
$ 0.813$ & $-0.054$ & 14 & E & 0.025 & HB\\
$ 0.657$ & $ 0.017$ & 14 & E & 0.041 & HB\\
$ 0.600$ & $ 0.072$ & 14 & E & 0.058 & HB\\
$ 0.506$ & $ 0.145$ & 14 & E & 0.077 & HB\\
$ 0.478$ & $ 0.286$ & 14 & E & 0.096 & HB\\
$ 0.494$ & $ 0.391$ & 14 & E & 0.114 & HB\\
$ 0.442$ & $ 0.545$ & 14 & E & 0.134 & HB\\
$ 0.063$ & $ 0.809$ & 14 & E & 0.162 & HB\\
$-0.620$ & $ 0.944$ & 14 & E & 0.214 & HB\\
$-2.554$ & $ 1.322$ & 1 & E & 0.236 & RGB/MS\\
$-2.509$ & $ 1.291$ & 1 & E & 0.258 & RGB/MS\\
$-2.233$ & $ 1.250$ & 1 & E & 0.274 & RGB/MS\\
$-2.198$ & $ 1.225$ & 1 & E & 0.290 & RGB/MS\\
$-2.138$ & $ 1.215$ & 1 & E & 0.305 & RGB/MS\\
$-2.023$ & $ 1.185$ & 1 & E & 0.319 & RGB/MS\\
$-1.821$ & $ 1.163$ & 1 & E & 0.330 & RGB/MS\\
$-1.480$ & $ 1.117$ & 1 & E & 0.338 & RGB/MS\\
$-1.393$ & $ 1.113$ & 1 & E & 0.346 & RGB/MS\\
$-1.384$ & $ 1.088$ & 1 & E & 0.353 & RGB/MS\\
$-1.336$ & $ 1.075$ & 1 & E & 0.361 & RGB/MS\\
$-1.323$ & $ 1.061$ & 1 & E & 0.368 & RGB/MS\\
$-1.283$ & $ 1.060$ & 1 & E & 0.375 & RGB/MS\\
$-1.144$ & $ 1.059$ & 1 & E & 0.381 & RGB/MS\\
$-1.108$ & $ 1.042$ & 1 & E & 0.387 & RGB/MS\\
$-1.087$ & $ 1.044$ & 1 & E & 0.392 & RGB/MS\\
$-1.028$ & $ 1.030$ & 1 & E & 0.398 & RGB/MS\\
$-0.994$ & $ 1.052$ & 1 & E & 0.403 & RGB/MS\\
$-0.981$ & $ 1.044$ & 1 & E & 0.408 & RGB/MS\\
$-0.910$ & $ 1.026$ & 1 & E & 0.413 & RGB/MS\\
$-0.736$ & $ 1.011$ & 5 & E & 0.434 & RGB/MS\\
$-0.568$ & $ 0.973$ & 5 & E & 0.452 & RGB/MS\\
$-0.010$ & $ 0.941$ & 5 & E & 0.462 & RGB/MS\\
$ 0.043$ & $ 0.928$ & 5 & E & 0.472 & RGB/MS\\
$ 0.071$ & $ 0.933$ & 5 & E & 0.482 & RGB/MS\\
$ 0.164$ & $ 0.920$ & 5 & E & 0.491 & RGB/MS\\
$ 0.234$ & $ 0.915$ & 5 & E & 0.500 & RGB/MS\\
$ 0.297$ & $ 0.920$ & 5 & E & 0.508 & RGB/MS\\
$ 0.381$ & $ 0.908$ & 5 & E & 0.515 & RGB/MS\\
$ 0.479$ & $ 0.897$ & 5 & E & 0.522 & RGB/MS\\
$ 0.580$ & $ 0.876$ & 5 & E & 0.528 & RGB/MS\\
$ 0.696$ & $ 0.897$ & 5 & E & 0.534 & RGB/MS\\
$ 0.776$ & $ 0.874$ & 5 & E & 0.539 & RGB/MS\\
$ 0.878$ & $ 0.875$ & 5 & E & 0.544 & RGB/MS\\
$ 0.937$ & $ 0.854$ & 5 & E & 0.548 & RGB/MS\\
$ 0.966$ & $ 0.879$ & 5 & E & 0.552 & RGB/MS\\
$ 1.014$ & $ 0.869$ & 5 & E & 0.557 & RGB/MS\\
$ 1.067$ & $ 0.837$ & 5 & E & 0.561 & RGB/MS\\
$ 1.144$ & $ 0.845$ & 5 & E & 0.564 & RGB/MS\\
$ 1.206$ & $ 0.844$ & 5 & E & 0.568 & RGB/MS\\
$ 1.347$ & $ 0.835$ & 2.16e+01 & S & 0.581 & RGB/MS\\
$ 1.567$ & $ 0.817$ & 3.26e+01 & S & 0.597 & RGB/MS\\
$ 2.165$ & $ 0.791$ & 9.80e+01 & S & 0.625 & RGB/MS\\
$ 2.515$ & $ 0.764$ & 7.54e+01 & S & 0.641 & RGB/MS\\
$ 2.791$ & $ 0.737$ & 8.34e+01 & S & 0.654 & RGB/MS\\
$ 2.980$ & $ 0.719$ & 5.37e+01 & S & 0.662 & RGB/MS\\
$ 3.150$ & $ 0.665$ & 8.47e+01 & S & 0.671 & RGB/MS\\
$ 3.264$ & $ 0.614$ & 6.74e+01 & S & 0.679 & RGB/MS\\
$ 3.336$ & $ 0.609$ & 5.82e+01 & S & 0.684 & RGB/MS\\
$ 3.397$ & $ 0.565$ & 6.49e+01 & S & 0.690 & RGB/MS\\
$ 3.463$ & $ 0.543$ & 7.51e+01 & S & 0.697 & RGB/MS\\
$ 3.529$ & $ 0.531$ & 7.46e+01 & S & 0.703 & RGB/MS\\
$ 3.658$ & $ 0.508$ & 2.99e+02 & S & 0.724 & RGB/MS\\
$ 3.839$ & $ 0.519$ & 3.78e+02 & S & 0.748 & RGB/MS\\
$ 4.014$ & $ 0.500$ & 3.96e+02 & S & 0.768 & RGB/MS\\
$ 4.167$ & $ 0.513$ & 3.62e+02 & S & 0.785 & RGB/MS\\
$ 4.291$ & $ 0.528$ & 3.52e+02 & S & 0.799 & RGB/MS\\
$ 4.422$ & $ 0.539$ & 3.13e+02 & S & 0.810 & RGB/MS\\
$ 4.528$ & $ 0.567$ & 2.82e+02 & S & 0.820 & RGB/MS\\
$ 4.631$ & $ 0.567$ & 2.78e+02 & S & 0.828 & RGB/MS\\
$ 4.723$ & $ 0.572$ & 2.75e+02 & S & 0.835 & RGB/MS\\
$ 4.813$ & $ 0.587$ & 2.64e+02 & S & 0.842 & RGB/MS\\
$ 4.904$ & $ 0.603$ & 3.40e+02 & S & 0.850 & RGB/MS\\
$ 4.984$ & $ 0.609$ & 3.24e+02 & S & 0.857 & RGB/MS\\
$ 5.202$ & $ 0.635$ & 9.85e+02 & S & 0.874 & ISO\\
$ 5.443$ & $ 0.668$ & 1.35e+03 & S & 0.893 & ISO\\
$ 5.703$ & $ 0.707$ & 1.67e+03 & S & 0.911 & ISO\\
$ 5.951$ & $ 0.748$ & 1.69e+03 & S & 0.926 & ISO\\
$ 6.170$ & $ 0.787$ & 1.55e+03 & S & 0.937 & ISO\\
$ 6.389$ & $ 0.826$ & 1.67e+03 & S & 0.947 & ISO\\
$ 6.591$ & $ 0.864$ & 1.67e+03 & S & 0.955 & ISO\\
$ 6.822$ & $ 0.909$ & 2.05e+03 & S & 0.963 & ISO\\
$ 7.067$ & $ 0.959$ & 2.30e+03 & S & 0.971 & ISO\\
$ 7.318$ & $ 1.012$ & 2.48e+03 & S & 0.977 & ISO\\
$ 7.582$ & $ 1.070$ & 2.77e+03 & S & 0.982 & ISO\\
$ 7.851$ & $ 1.130$ & 3.17e+03 & S & 0.987 & ISO\\
$ 8.052$ & $ 1.173$ & 2.56e+03 & S & 0.990 & ISO\\
$ 8.264$ & $ 1.217$ & 3.02e+03 & S & 0.993 & ISO\\
$ 8.472$ & $ 1.257$ & 3.45e+03 & S & 0.996 & ISO\\
$ 8.735$ & $ 1.302$ & 5.39e+03 & S & 1.000 & ISO\\
\hline
\end{longtable}
}

Tables~\ref{tab:cmdf3}-\ref{tab:cmdf5} (available in the on-line edition only) list our final adopted CMDs for each cluster. The first two columns list the $M_V$ magnitude and $V\!-\!I$ colour for each cmd-bin, while the following column lists the weight of the corresponding bin when computing the integrated spectra. We used the photometry for the brightest 20 RGB stars directly, and the corresponding weight for each bin is then simply one. Elsewhere in the CMDs, larger numbers of stars were averaged together per bin, ranging from 5 stars per bin just below the 20 brightest individual stars to 100--150 stars per bin at the faint end. Each CMD then consists of about 100 bins. An ``E'' flag after the weight indicates that the weight is simply the number of observed stars per bin. An ``S'' flag instead indicates that the weights were determined from the model LFs, normalized to the empirical ones over the range $0 > M_V > -2$. The 5th column indicates the cumulative $V$-band flux for each bin. It can be seen that the HB contributes $\sim$5\%--20\% of the $V$-band flux, and a significant fraction of the flux comes from stars below the observational limit. Finally, the last column indicates the nature of the stars in each bin. Here, ``ISO'' indicates that the colours come from the \citet{Dotter2007} isochrones, while ``RIDGE'' (in the case of Fornax 4) is the ridge line from \citet{Buonanno1999}.

The binned CMDs are shown in Fig.~\ref{fig:cmds}. Note the smooth continuation from empirical to theoretical CMDs somewhat below the main sequence turn-off. For Fornax 4 we used the empirical ridge line from \citet{Buonanno1999} in the range $2 < M_V < 4$ where the contamination of the CMD from the general field becomes  significant.
 We have drawn the theoretical isochrones from \citet{Dotter2007} on top of each CMD. They fit the observed CMDs fairly well, although some slight mismatches are seen at the faint end of the RGB.

\section{Abundance analysis from integrated spectra}
\label{sec:analysis}

We now explain in more detail our approach to measuring detailed chemical abundances from the integrated-light spectra.  In essence, we simply calculate a series of SSP models at high spectral resolution, taking into account all stars across the H-R diagram and making use of model atmospheres and spectral synthesis to compute the individual model spectra for a given chemical composition.
We calculate these models ``on-the-fly'' while iteratively solving for the best-fitting abundances, i.e., we do not make use of pre-computed spectral libraries. 
Naturally,  some trade-offs are involved.
To make this computationally feasible, we rely (at least for the time being) on classical model atmospheres and synthetic spectral calculations, specifically those computed with the publicly available \texttt{ATLAS9} \citep{Kurucz1970} and \texttt{SYNTHE} \citep{Kurucz1979,Kurucz1981} codes. We use the versions of these codes available as of April 2011 from the website of F.\ Castelli\footnote{http://wwwuser.oat.ts.astro.it/castelli/}  \citep{Sbordone2004}, from which we also obtained ancillary data (line lists, molecular data, opacity distribution functions, etc.). 
The Kurucz codes were compiled with the Intel Fortran compiler for Mac OS X.
The Kurucz models are based on the common assumption of Local Thermodynamic Equilibrium (LTE), as well as being one-dimensional, static and plane parallel. Although none of these approximations are ever completely valid, they simplify the calculations enormously and are often made in the analysis of globular cluster stars. 
In younger clusters, where stars with higher temperatures and/or low surface gravities contribute a significant fraction of the light, non-LTE effects will likely start to play a more significant role.
The limitations associated with the use of classical model atmospheres have been thoroughly discussed by \citet{Asplund2005}.
The line lists are based on those available from the Kurucz web site\footnote{http://kurucz.harvard.edu/linelists.html}, with some modifications \citep{Castelli2004}. They include essentially all atomic transitions with laboratory data, as well as several diatomic molecules (CH, MgH, NH, OH, SiH, H$_2$, C$_2$, CN, CO, SiO and TiO) and H$_2$O. Of the elements analysed here, hyperfine splitting is included for \ion{Sc}{ii} ($\lambda 5527, 5658$ \AA), \ion{Mn}{ii} ($\lambda 4756, 4765, 4783, 4824$ \AA) and \ion{Ba}{ii} ($\lambda 4554, 4934, 5854, 6142$ \AA).
The line list includes damping constants for radiative, Stark and van der Waals damping, along with literature references for these constants. When literature values are not available, they are computed internally in \texttt{SYNTHE} using an approximate expression \citep{Kurucz1981}.

\subsection{General overview}

In detail, then, we proceeded as follows: We first used the CMDs in Fig.~\ref{fig:cmds} to derive physical parameters for each cmd-bin (where a ``cmd-bin'' may refer to a single star or a bin containing  several stars). From the mean $V\!-\!I$ colours of each cmd-bin we derived effective temperatures $T_\mathrm{eff}$ and bolometric corrections $BC$, using the Kurucz colour-$T_\mathrm{eff}$ transformations.  Surface gravities, $\log g$,  then followed from the standard relation
\begin{equation}
  \log g = \log g_\odot + \log \left[\left(\frac{T_\mathrm{eff}}{T_\mathrm{eff,\odot}}\right)^4 \left(\frac{M}{M_\odot}\right) \left(\frac{L_\mathrm{bol}}{L_\mathrm{bol,\odot}}\right)^{-1}\right]
  \label{eq:logg}
\end{equation} 
Here we assumed masses of 0.9 $M_\odot$ for the empirical data, corresponding roughly to the main sequence turn-off mass for a 10 Gyr-old population, while the masses tabulated in the isochrones were used for the theoretical data. At the level of precision to which the $\log g$ values are required, this is sufficiently accurate.
Because the transformations from colour to $T_\mathrm{eff}$ and $BC$ depend (weakly) on $\log g$, we iterated a few times on Eq.~(\ref{eq:logg}). As the transformations also depend on metallicity, we started the analysis with  initial ``guesses'' based on the RGB slope \citep{Buonanno1998,Buonanno1999}, but then self-consistently used our own metallicity determinations.

Having established the physical parameters, we computed an \texttt{ATLAS9} model atmosphere for each cmd-bin. We used the NEWODF opacity distribution functions \citep[ODFs;][]{Castelli2003} with solar-scaled abundances, but  verified that the results did not change significantly for $\alpha$-enhanced ODFs (Sect.~\ref{sec:uncertainties}).
We then computed synthetic spectra with \texttt{SYNTHE}. In order to speed up this procedure, the TiO lines were skipped for stars with $T_\mathrm{eff} > 4500$ K where the TiO features vanish anyway \citep[e.g.][]{Mould1978}. The spectra were initially computed at very high resolution ($R = 500\,000$).
The synthetic spectra were then scaled to the stellar luminosity corresponding to each cmd-bin, multiplied by the weight of the cmd-bin, and finally co-added to a single integrated spectrum.
The integrated model spectra were broadened to match the data, and the ratio of the continua of the model- and synthetic spectra were fitted with a 3rd order spline function or a polynomial, depending on the wavelength range fitted. The broadened model spectra were then multiplied by the spline/polynomial fit, subtracted from the observed spectra, and the $\chi^2$ evaluated. We include an option to assign different weights to different parts of the spectra on a pixel-to-pixel basis. In practice, only weights of 1.0 (good) or 0.0 (bad) were used but intermediate values are in principle also allowed.
Note that we make no formal distinction between ``continuum'' or ``feature'' regions for the derivation of abundances; the entire spectrum is fitted. 
Prior to this analysis, the observed spectra were shifted to zero radial velocity by comparison with a model spectrum. 

The best-fitting abundances were determined by iteratively searching for the abundances that minimized the $\chi^2$. In practice, this was implemented by calling  \texttt{ATLAS9} and \texttt{SYNTHE} from a ``wrapper'' code written in the \texttt{Python} language. 
This code allows us to fit for arbitrary combinations of individual abundances simultaneously.  Elements can be forced to vary in lockstep (e.g., the $\alpha$-elements), while others may be kept fixed at user-specified abundances (e.g. from a previous run). The overall scaling of all abundances (relative to Solar) can be kept fixed, or allowed to vary as a free parameter during the fit. The smoothing and microturbulence can also be kept fixed, or allowed to vary as free parameters. 
When multiple parameters are fit simultaneously we use the ``downhill simplex'' algorithm \citep{Nelder1965,Press1992}, while a ``golden section'' search is used when only one abundance is being fit. In the latter case, the code can also output one-sigma error intervals (by varying the abundance until $\chi^2 = \chi^2_\mathrm{min}+1$), based on the errors provided for the observed spectrum. Abundances quoted in this paper are given relative to the Solar composition \citep{Grevesse1998}.

The model atmospheres were only computed once, at the beginning of each run. Computing new atmospheres each iteration would significantly increase the computational burden, and we instead leave it up to the user to provide reasonable initial guesses for the chemical composition of the stars. If these are poorly known, it may be necessary to iterate once or twice.  
To make efficient use of modern multiple-core computers, we have set up the \texttt{Python} code to run multiple instances of \texttt{ATLAS9} and \texttt{SYNTHE} in parallel. The analysis in this paper was carried out on a 12-core Mac Pro workstation. The time required per fit depended on many factors, ranging from  a few minutes (if only a single abundance was measured over a small wavelength range, and no errors were required) to a few hours (fitting multiple abundances over a 200 \AA\ region).
We note that while we are currently relying on the Kurucz codes as a ``backbone'', most of the \texttt{Python} code is not specific to \texttt{ATLAS9} and \texttt{SYNTHE} and implementing other model atmospheres and/or spectral synthesis codes should be relatively straight forward. In particular, the code currently contains  ``hooks'' for the \texttt{MOOG} code \citep{Sneden1973} although this is not yet fully implemented.

\subsection{Microturbulence}
\label{sec:vt}

The microturbulent velocity $v_t$ can be fitted as a free parameter (which is then, in the current implementation, common to all stars) or specified individually for each bin in the input CMD. We carried out a set of test runs in which we fitted for the overall scaling of the abundances, the Fe and $\alpha$-element abundances, and the microturbulence, in adjacent 200~\AA\  bins from 4200~\AA\ -- 6200~\AA.  For the more metal-poor clusters (Fornax~3 and Fornax~5) we found $v_t$ to be poorly constrained by the data, with large bin-to-bin dispersion. This is not surprising, given that the equivalent widths of the mostly weak lines in these spectra are relatively insensitive to the amount of microturbulence. For Fornax~4, we instead found the microturbulence to be well constrained with a mean $\langle v_t \rangle = 2.0$ km s$^{-1}$ with a bin-to-bin dispersion of $\sigma(v_t) = 0.18$ km s$^{-1}$.
The use of a single $v_t$ value for all stars in a cluster is probably an oversimplification, however. While our best-fit value for Fornax~4 is typical of the values determined for individual bright RGB stars in globular clusters \citep{Kraft1992,Pilachowski1996} including the Fornax GCs \citep{Letarte2006}, $v_t$ tends to decrease with increasing surface gravity. For subgiants and stars near the turn-off, a typical value is $v_t\sim1$ km s$^{-1}$ \citep{Carretta2004}. \citet{McWilliam2008} parameterized this variation as a linear function of $\log g$, constrained by the $\log g$ and $v_t$ values for Arcturus and the Sun.
For our purpose, it was more convenient to parameterize the trend as a function of direct observables, since this allowed us to assign the $v_t$ values to each cmd-bin in a transparent and straight forward way, prior to further analysis. A plot of our derived $\log g$ values vs. $M_V$ showed these two quantities to be well approximated by a linear relation for $M_V < +4$. We therefore adopted two reference points: $(\log g, v_t)_1$ = (1.0, 2.0 km s$^{-1}$) and $(\log g, v_t)_2$ = (4.0, 1.0 km s$^{-1}$), corresponding to $(M_V, v_t)_1$ = ($-1.81$ mag, 2.0 km s$^{-1}$) and $(M_V, v_t)_2$ = ($+3.64$ mag, 1.0 km s$^{-1}$).
For cmd-bins brighter than $(M_V)_2$ that represented non-HB stars, we assigned $v_t$ values based on a linear fit to these two points. For cmd-bins fainter than $(M_V)_2$ we assumed a constant $v_t = 1$ km s$^{-1}$, and for cmd-bins representing horizontal branch stars we assumed $v_t = 1.8$ km s$^{-1}$ \citep{Pilachowski1996}.

\subsection{Smoothing of the model spectra}

\begin{figure}
\includegraphics[width=\columnwidth]{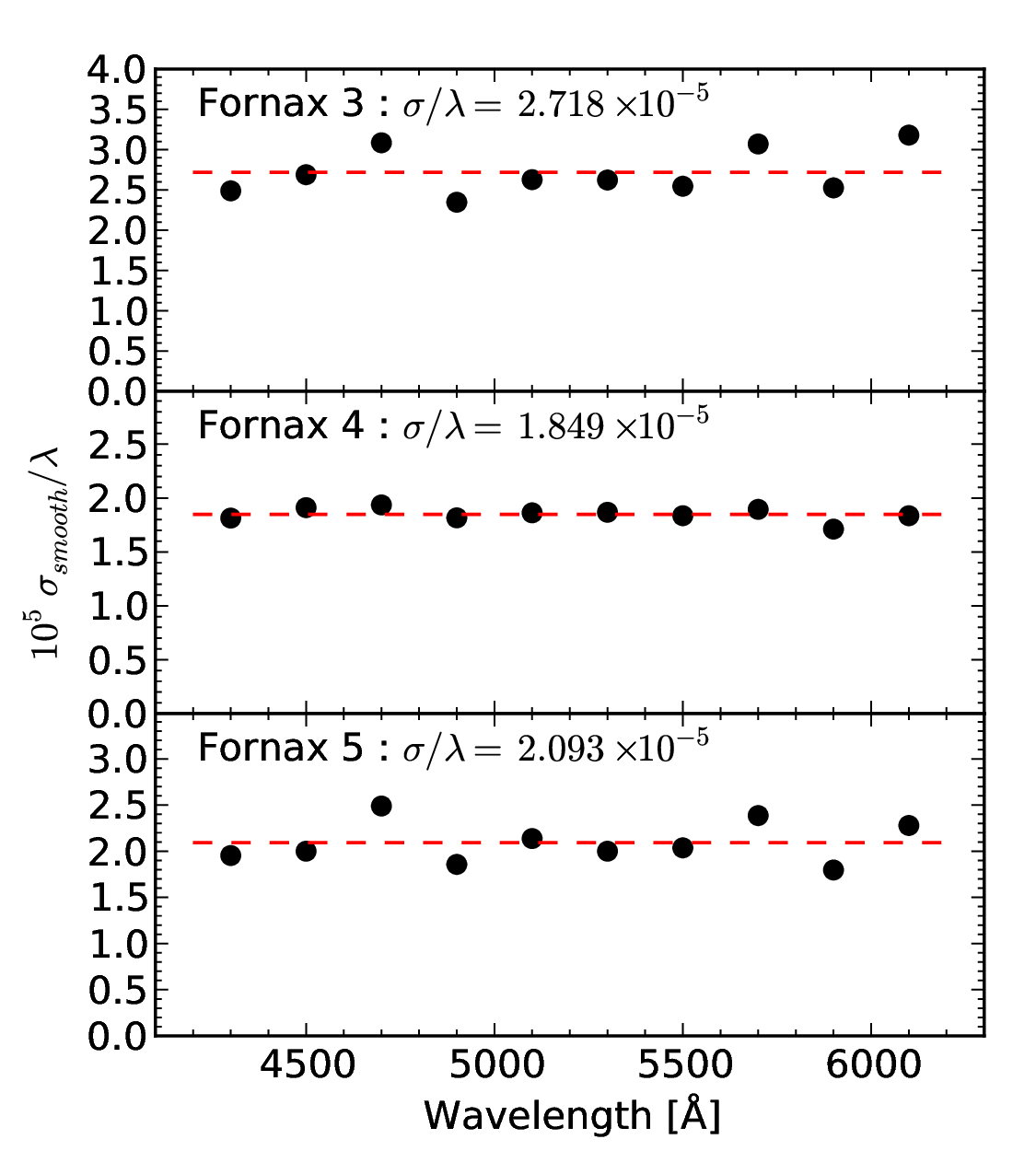}
\caption{\label{fig:sigfit}The ratio of smoothing scale vs.\ wavelength, $\sigma_\mathrm{smooth}/\lambda$, as a function of wavelength. The mean $\langle\sigma_\mathrm{smooth}/\lambda\rangle$ is  indicated by a dashed line.}
\end{figure}

The model spectra must be degraded to fit the data. This is mainly due to the finite instrumental resolution and internal velocity dispersions in the clusters, although other effects also contribute to the broadening of spectral features (e.g.\ stellar rotation, macroturbulence, etc.). These effects are all expected to scale as $\sigma_\mathrm{smooth} / \lambda \simeq$ const ($\sigma_\mathrm{smooth}$ being the dispersion in wavelength units of the Gaussian kernel used for the smoothing). This is clearly the case for the various velocity-related broadenings, but, in our case, also (roughly) for the instrumental broadening which is dominated by the slit width: Ignoring distortions by the instrument optics, the projected slit image has a roughly constant width in pixel units, $s_p$, set mainly by the focal length of the camera. From the grating equation,
the dispersion scales with  echelle order $m$ as $dx/d\lambda\propto m$, so that the width of the slit will also be roughly constant  in  wavelength units ($s_\lambda$) \emph{within} a given echelle order. However, since the product $m\lambda \sim$ const across the different echelle orders, we expect the ratio $s_\lambda / \lambda \simeq$ const over a wide wavelength range, as well. Not accounting for other effects, the instrumental resolution $s_\lambda/\lambda$ should thus be roughly constant, modulated by a trend with wavelength within each echelle order. This latter effect is small, however, because the UVES echelle grating is used in very high orders ($m = 100-145$) so that the intra-order variation in the ratio $s_\lambda/\lambda$ is only expected to be $\sim$1\%. We therefore ignored this more complicated behaviour, especially since the instrumental broadening turned out to be relatively small compared to that introduced by the relative motions of the stars in the clusters, as discussed below.

While our fitting procedure allowed us to solve for $\sigma_\mathrm{smooth}$ as a free parameter, this is not generally desirable as it introduces an extra degree of freedom in the fits that in some cases may dominate the uncertainties on the derived abundances. We therefore carried out a new set of fits to each spectrum in 200~\AA\ bins, now fixing the microturbulence $v_t$ to the relation discussed in Sect.~\ref{sec:vt}, but still allowing $\sigma_\mathrm{smooth}$ and the Fe and $\alpha$-element abundances to vary. These runs then provided the smoothing and initial estimates of chemical composition used in subsequent analysis.
Figure~\ref{fig:sigfit} shows the resulting $\sigma_\mathrm{smooth}/\lambda$ vs.\ $\lambda$ relations for each cluster. While $\sigma_\mathrm{smooth}/\lambda$ is roughly constant with $\lambda$, as expected, the residuals in each bin do display significant systematics when comparing the different clusters (some bins are systematically high, others systematically low). This serves to emphasize that significant systematic errors might occur if $\sigma_\mathrm{smooth}$ were allowed to vary freely.

From this initial set of fits, we also obtained $\mathrm{[Fe/H]} = -2.35$ (Fornax~3), 
$\mathrm{[Fe/H]} = -1.40$ (Fornax~4) and $\mathrm{[Fe/H]} = -2.06$ (Fornax~5). The bin-to-bin scatter is $\sigma_{\rm [Fe/H]} = 0.12$ dex, $0.06$ dex and $0.16$ dex for the three clusters, respectively. These fits also returned mean [$\alpha$/Fe] ratios of [$\alpha$/Fe] = $+0.12$ (Fornax~3), [$\alpha$/Fe] = $-0.02$ (Fornax~4) and [$\alpha$/Fe] = $0.14$ (Fornax~5) with bin-to-bin scatter of $\sigma_{\rm [\alpha/Fe]} =$ 0.11 dex, 0.07 dex and 0.18 dex. The smaller bin-to-bin scatter for Fornax~4 is presumably due to the stronger  features in the spectrum of this cluster.
While these numbers are subject to further refinement and more careful analysis in the following sections, we already note that our [Fe/H] measurement for Fornax~3 agrees reassuringly well with the value of [Fe/H] = $-2.4\pm0.1$ derived by \citet{Letarte2006} from high-dispersion spectroscopy of individual stars. It is also clear that Fornax~4 is significantly more metal-rich than Fornax~3 and Fornax~5,
and may have a somewhat lower [$\alpha$/Fe] ratio than the other two clusters.

\section{Results and uncertainties}
\label{sec:results}

\begin{table*}
\caption{Individual abundance measurements.}
\begin{minipage}{\columnwidth}
\label{tab:rawabun}
\renewcommand{\footnoterule}{}
\begin{tabular}{lccc} \hline\hline
             $\lambda$ (\AA) & Fornax 3 & Fornax 4 & Fornax 5 \\ \hline
$[$Fe/H$]$ \\
4200--4400 & $-2.398\pm0.016$ & $-1.487\pm0.006$ & $-2.063\pm0.016$ \\
4400--4600 & $-2.389\pm0.016$ & $-1.455\pm0.011$ & $-2.163\pm0.016$ \\
4600--4800 & $-2.361\pm0.016$ & $-1.407\pm0.011$ & $-2.097\pm0.025$ \\
4800--5000 & $-2.392\pm0.011$ & $-1.439\pm0.011$ & $-2.195\pm0.016$ \\
5000--5170 & $-2.255\pm0.011$ & $-1.399\pm0.011$ & $-2.142\pm0.016$ \\
5230--5400 & $-2.268\pm0.011$ & $-1.372\pm0.011$ & $-2.037\pm0.021$ \\
5400--5600 & $-2.311\pm0.016$ & $-1.372\pm0.011$ & $-2.184\pm0.021$ \\
5600--5800 & $-2.422\pm0.025$ & $-1.345\pm0.011$ & $-2.023\pm0.036$ \\
5800--6000 & $-2.457\pm0.100$ & $-1.352\pm0.016$ & $-1.947\pm0.056$ \\
6000--6200 & $-2.170\pm0.031$ & $-1.285\pm0.016$ & $-1.711\pm0.021$ \\
$[$Mg/Fe$]$ \\
4352             & $+0.137\pm0.086$ & $-0.344\pm0.086$  & $+0.526\pm0.131$ \\
4571             & $+0.076\pm0.086$ & $-0.044\pm0.066$  & $+0.307\pm0.115$ \\
4703             & $-0.195\pm0.056$  & $+0.015\pm0.041$ & $-0.034\pm0.081$ \\
5167             & $-0.195\pm0.076$  & $-0.115\pm0.051$  & $+0.137\pm0.111$ \\
5528             & $-0.034\pm0.061$  & $+0.046\pm0.056$ & $-0.034\pm0.096$ \\
5711             & $+0.395\pm0.110$ & $-0.024\pm0.051$  & $+0.297\pm0.151$ \\
$[$Ca/Fe$]$ \\
4227             & $+0.166\pm0.040$ & $-0.003\pm0.031$  & $+0.256\pm0.061$ \\
4280--4320 & $+0.147\pm0.036$ & $+0.236\pm0.041$ & $+0.376\pm0.056$ \\
4420--4460 & $+0.266\pm0.025$ & $+0.116\pm0.025$ & $+0.147\pm0.051$ \\
4575--4591 & $-0.134\pm0.096$  & $-0.014\pm0.041$  & $-0.244\pm0.131$ \\
4878             & $+0.616\pm0.076$ & $+0.567\pm0.070$ & $+0.607\pm0.091$ \\
5259--5268 & $+0.366\pm0.066$ & $+0.256\pm0.041$ & $+0.395\pm0.086$ \\
5580--5610 & $+0.116\pm0.041$ & $+0.207\pm0.026$ & $+0.307\pm0.046$ \\
6100--6175 & $+0.326\pm0.025$ & $+0.105\pm0.016$ & $+0.236\pm0.03$ \\
$[$Sc/Fe$]$ \\
4290--4330 & $+0.046\pm0.046$ & $+0.127\pm0.046$ & $+0.026\pm0.076$ \\
4350--4440 & $+0.297\pm0.036$ & $-0.034\pm0.036$  & $+0.046\pm0.061$ \\
4670             & $-0.454\pm0.156$  & $-0.364\pm0.086$  & $-0.383\pm0.156$ \\
5031             & $-0.183\pm0.091$  & $-0.034\pm0.046$  & $-0.723\pm0.150$ \\
5527             & $+0.276\pm0.086$ & $-0.124\pm0.070$  & $-0.183\pm0.166$ \\
5638--5690 & $+0.256\pm0.051$ & $-0.064\pm0.030$  & $+0.105\pm0.081$ \\ \hline
\end{tabular}
\end{minipage}
\begin{minipage}{\columnwidth}
\begin{tabular}{lccc} \hline\hline
             $\lambda$ (\AA) & Fornax 3 & Fornax 4 & Fornax 5 \\ \hline
$[$Ti/Fe$]$ \\
4292--4320 & $+0.266\pm0.030$ & $-0.003\pm0.030$  & $+0.195\pm0.056$ \\
4386--4420 & $+0.416\pm0.025$ & $+0.286\pm0.021$ & $+0.307\pm0.046$ \\
4440--4474 & $+0.315\pm0.026$ & $+0.095\pm0.021$ & $+0.195\pm0.051$ \\
4532--4574 & $+0.217\pm0.021$ & $+0.066\pm0.021$ & $+0.307\pm0.036$ \\
4590             & $+0.476\pm0.075$ & $+0.447\pm0.066$ & $+0.567\pm0.111$ \\
4650--4715 & $+0.276\pm0.046$ & $+0.026\pm0.025$ & $+0.166\pm0.056$ \\
4750--4850 & $+0.447\pm0.026$ & $+0.207\pm0.021$ & $+0.386\pm0.046$ \\
4980--5045 & $+0.156\pm0.021$ & $+0.095\pm0.011$ & $+0.137\pm0.031$ \\
5154             & $+0.217\pm0.076$ & $+0.276\pm0.051$ & $+0.195\pm0.096$ \\
$[$Cr/Fe$]$ \\
4252--4292 & $-0.154\pm0.041$ & $-0.014\pm0.031$  & $+0.346\pm0.066$ \\
4353--4400 & $-0.214\pm0.070$ & $-0.364\pm0.051$  & $-0.144\pm0.096$ \\
4520--4660 & $-0.205\pm0.035$ & $-0.073\pm0.016$  & $-0.044\pm0.041$ \\
5235--5330 & $-0.244\pm0.061$ & $-0.034\pm0.025$  & $-0.093\pm0.060$ \\
5342--5351 & $+0.026\pm0.076$ & $+0.127\pm0.051$ & $+0.307\pm0.101$ \\
5410             & $+0.005\pm0.086$ & $+0.076\pm0.070$ & $-0.324\pm0.146$ \\
$[$Mn/Fe$]$ \\
4750--4790 & $-0.523\pm0.070$  & $-0.263\pm0.025$ & $-0.413\pm0.070$ \\
$[$Cu/Fe$]$ \\
5106             & $<-1$      & $-0.834\pm0.076$ & $-0.555\pm0.210$ \\
$[$Y/Fe$]$ \\
4355--4425 & $-0.173\pm0.071$ & $-0.289\pm0.065$ & $-0.310\pm0.121$ \\
4884             & $-0.115\pm0.116$ & $-0.334\pm0.076$ & $-0.164\pm0.171$ \\
5088             & $+0.266\pm0.086$ & $-0.134\pm0.070$ & $-0.243\pm0.146$ \\
$[$Ba/Fe$]$ \\
4554             & $+0.856\pm0.035$ & $+0.366\pm0.030$ & $+0.166\pm0.080$ \\
4934             & $+0.997\pm0.056$ & $+0.806\pm0.041$ & $+0.485\pm0.075$ \\
5846             & $+0.376\pm0.086$ & $+0.185\pm0.081$ & $-0.173\pm0.150$ \\
6142             & $+0.707\pm0.056$ & $+0.227\pm0.051$ & $-0.205\pm0.121$ \\
$[$Eu/Fe$]$ \\
4205             & $+1.527\pm0.131$ & $+2.197\pm0.096$ & $+1.286\pm0.206$ \\
4436             & $+1.477\pm0.126$ & $+0.606\pm0.146$ & $+0.177\pm0.405$ \\
\hline
\end{tabular}
\vspace{3.2mm}
\end{minipage}
\tablefoot{
A wavelength range in the first column indicates that this range contains several absorption features from the corresponding element. When a single wavelength is given, a single line is fitted within a 10~\AA\ window.
}
\end{table*}

\begin{table*}
\begin{minipage}[t]{17cm}
\caption{Average integrated-light abundance measurements. 
}
\label{tab:abun}
\renewcommand{\footnoterule}{}
\begin{tabular}{lcrrrrrr} \hline\hline
            & $N$ & \multicolumn{2}{c}{Fornax 3} & \multicolumn{2}{c}{Fornax 4} & \multicolumn{2}{c}{Fornax 5} \\ 
            &          & weighted avg & r.m.s. &  weighted avg & r.m.s. & weighted avg & r.m.s. \\   \hline
$\sigma_{\rm vd}$ [km s$^{-1}$] & 10 & 7.5 & 0.8 & 4.5 & 0.2 &  5.4 & 0.6  \\ 
$[$Fe/H$]$   & 10 & $-2.33\pm0.01$ & 0.09 &  $-1.42\pm0.01$ & 0.06 & $-2.09\pm0.01$ & 0.14   \\   
$[$Mg/Fe$]$ & 6   & $-0.04\pm0.03$ & 0.22 &  $-0.04\pm0.02$  & 0.13 & $+0.13\pm0.04$ & 0.21 \\   
$[$Ca/Fe$]$ &  8  & $+0.25\pm0.01$ & 0.20 &  $+0.13\pm0.01$ & 0.18& $+0.27\pm0.02$ & 0.23 \\   
$[$Sc/Fe$]$  &  6  & $+0.18\pm0.02$ & 0.31 &  $-0.04\pm0.02$ & 0.15 &$-0.03\pm0.04$ & 0.33 \\     
$[$Ti/Fe$]$   & 9   & $+0.29\pm0.01$ & 0.11 & $+0.12\pm0.01$ & 0.15 & $+0.24\pm0.02$ & 0.13 \\  
$[$Cr/Fe$]$  & 6   & $-0.17\pm0.02$  & 0.11 & $-0.06\pm0.01$  & 0.16 & $+0.02\pm0.03$ & 0.24 \\   
$[$Mn/Fe$]$ & 1   & $-0.52\pm0.07$  & -       &  $-0.26\pm0.03$ & -         & $-0.41\pm0.07$ & -  \\     
$[$Cu/Fe$]$ & 1   & $<-1.0 \, (2\sigma)$ & - & $-0.83\pm0.08$ & - & $-0.56\pm0.21$ & - \\   
$[$Y/Fe$]$    & 3   & $-0.02\pm0.04$  & 0.19 & $-0.25\pm0.04$ & 0.09  & $-0.26\pm0.08$ & 0.06 \\        
$[$Ba/Fe$]$  & 4   &  $+0.81\pm0.02$& 0.24 & $+0.45\pm0.02$ & 0.25 & $+0.20\pm0.05$ & 0.31 \\  
$[$Eu/Fe$]$  & 2   & $+1.50\pm0.09$ & 0.03 & $+1.72\pm0.08$ & 0.86 & $+1.06\pm0.18$ & 0.64 \\   
\hline
\end{tabular}
\end{minipage}
\tablefoot{
$N$ is the number of individual fits for each element (see Table~\ref{tab:rawabun}). For each entry we give the weighted average of the measurements in Table~\ref{tab:rawabun} and the r.m.s. scatter.
}
\end{table*}

\begin{table*}
\begin{minipage}[t]{130mm}
\caption{Systematic uncertainties on abundance measurements. 
}
\label{tab:systematics}
\renewcommand{\footnoterule}{}
\begin{tabular}{lrrrrrrrrr} \hline\hline
            &  \multicolumn{3}{c}{$\Delta E(V\!-\!I) = 0.05$} & \multicolumn{3}{c}{$\Delta v_t = 0.5$ km s$^{-1}$} & \multicolumn{3}{c}{$M_{V,\rm lim}$ 9$\rightarrow$13 mag$^a$} \\
            & F3 & F4 & F5 & F3 & F4 & F5 & F3 & F4 & F5 \\ \hline
$\Delta [$Fe/H$]$   & $+$0.09 & $+$0.09 & $+$0.13 & $-$0.11 & $-$0.11 & $-$0.12 & $-$0.07 & $-$0.01 & $-$0.04 \\
$\Delta [$Mg/Fe$]$ & $-$0.04 & 0.00 & $-$0.04 & $+$0.06 & $+$0.01 & $+$0.03 & $-$0.03 & $-$0.05 & $-$0.04 \\
$\Delta [$Ca/Fe$]$ &  0.00      & $+$0.01 & $-$0.02 & $+$0.03 & $-$0.04 & $+$0.03 & $-$0.10 & $-$0.10 & $-$0.10 \\
$\Delta [$Sc/Fe$]$  & $-$0.02 & $-$0.05 & $-$0.01 & $+$0.01 & $-$0.07 & $+$0.00 & $+$0.09 & $+$0.03 & $+$0.05 \\
$\Delta [$Ti/Fe$]$   & $-$0.02 & $-$0.02 & $+$0.03 &  $-$0.01 & $-$0.06 & $-$0.01 & $+$0.00 & $-$0.04 & $-$0.02 \\
$\Delta [$Cr/Fe$]$  & $+$0.02 & 0.00 & 0.00 & $+$0.02 & $-$0.02 & $+$0.02 & $-$0.11 & $-$0.05 & $-$0.06 \\
$\Delta [$Mn/Fe$]$ & 0.00 & 0.00 & $-$0.01 & $+0.08$ & $-0.01$ & $+$0.07 & $-$0.01 & $-$0.01 & $-$0.02 \\
$\Delta [$Cu/Fe$]$ & $-$ & $+$0.02 & $+$0.01 & $-$ & $+$0.06 & $+$0.10 & $-$ & $+$0.02 & $+$0.06 \\
$\Delta [$Y/Fe$]$    & 0.00 & $-$0.06 & 0.00 & $+$0.02 & $-$0.10 & $+$0.01 & $+$0.17 & $+$0.06 & $+$0.09 \\
$\Delta [$Ba/Fe$]$  & $+$0.02 & $-$0.01 & $+$0.01 & $-$0.11 & $-$0.13 & $-$0.11 & $-$0.00 & $-$0.02 & $-$0.00\\
$\Delta [$Eu/Fe$]$  & $-$0.04 & $-$0.04 & $-$0.03 & $-$0.01 & $-$0.23 & $+$0.01& $+$0.02 & $-$0.05 & $-$0.05 \\
\hline
\end{tabular}
\end{minipage}
\tablefoot{
\tablefoottext{a}{Indicates a change in the limiting magnitude of the CMD modelling from $M_V=+9$ mag to $M_V=+13$ mag.}
}
\end{table*}

\begin{figure*}
\centering
\includegraphics[width=180mm]{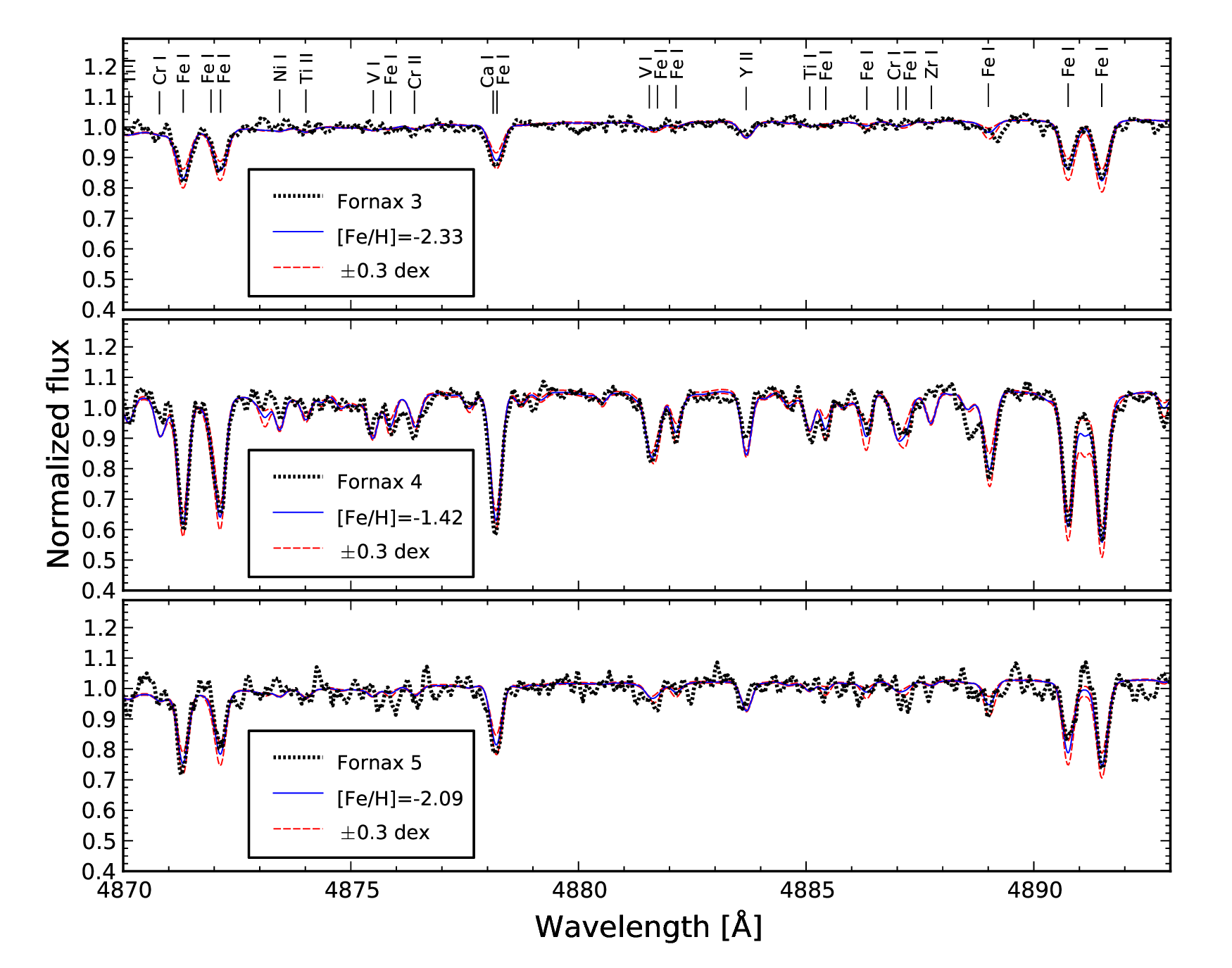}
\caption{\label{fig:fitfe}Fits to a small part of the UVES spectra (dotted, thick black curves). The thin, solid curves (blue in the on-line edition) show the synthetic spectra for the overall best-fitting Fe abundances, while the thin dashed curves (red in the on-line edition) show the effect of of varying the Fe abundance by $\pm0.3$ dex. The spectra have been smoothed with a 5 pixels wide boxcar filter. }
\end{figure*}

Our initial fits reported in Sect.~\ref{sec:analysis} were carried out in fixed 200~\AA\ wavelength intervals. However, the measurable features of individual elements are not uniformly distributed across the spectra, and for our more detailed abundance analysis we concentrated on a number of selected wavelength windows. These were defined by visually inspecting the spectra with individual absorption features labelled. Although the higher metallicity of Fornax~4 in principle allowed the measurement of a larger number of weaker lines, we kept the same windows for all three clusters for consistency reasons.
While our procedure allowed us to fit for multiple abundances simultaneously, in practice it sufficed to fit one element at the time in most cases. 

We first fit for the Fe abundances. Because of the large number of Fe lines found throughout the spectra, we simply divided the spectra into 200~\AA\ bins as before and fit for the Fe abundance in each of the resulting 10 bins while keeping the smoothing in each bin fixed according to the relations in Fig.~\ref{fig:sigfit}. 
The Fe abundances  were obtained by fitting not only for Fe, but also for other elements that contained prominent absorption lines in the respective bins. However, for most bins it made very little difference whether we actually let these other abundances vary or kept them fixed. 
The notable exception was the 4200~\AA\ -- 4400~\AA\ range which contains prominent CH absorption bands; if the C abundance was forced to solar-scaled values this generally led to significantly higher [Fe/H] values in this wavelength range (by up to 0.4 dex in the most extreme case of Fornax 4) compared to the other bins. By letting the C abundance vary, the [Fe/H] abundance in the 4200~\AA\ -- 4400~\AA\ range was instead consistent with those obtained at other wavelengths. 
From the fits to the 4200~\AA\ -- 4400~\AA\ and 4400~\AA\ -- 4600~\AA\ bins, we found C to be significantly depleted compared to Fe, with  [C/Fe] ratios in the range $-0.5$ to $-0.7$. 
It is well known that the $[$C/Fe$]$ ratio decreases as a function of luminosity for RGB stars, particularly at low metallicities where C can be depleted by nearly 1 dex for the brightest RGB stars \citep{Martell2008}. In our integrated-light spectra we only measured the luminosity-weighted mean $[$C/Fe$]$ values, which are clearly significantly sub-solar. However, it should be kept in mind that these values were obtained by scaling the C abundance of all stars in the H-R diagrams, and as such our integrated synthetic spectra should be considered less reliable in the 4200~\AA\ -- 4400~\AA\ region where the CH features are strong. It should, in principle,  be possible to allow for a (suitably parameterized) variation in the $[$C/Fe$]$ ratio along the RGB. 
To be fully realistic, such modelling would  also need to account for variations in the carbon isotope ratios as a function of stellar type. The Kurucz code uses a hard-coded ratio of \element[][12]{C}/\element[][13]{C} = 89, i.e. the Solar value \citep[e.g.][]{Harris1987}, while giant stars generally have much lower isotopic ratios of
\element[][12]{C}/\element[][13]{C} $\la 10$ \citep{Gratton2000,Keller2001}.
The integrated-light measurements of C likely reflect a combination of effects related to both stellar evolution and the overall chemistry of the clusters,
and are clearly complicated to interpret.
For these reasons, we do not include C in our further discussion of individual elements below.

When measuring  the remaining elements, we kept the overall scalings of the abundances fixed according to the derived [Fe/H] and [C/Fe] abundances. We then proceeded by first fitting the elements with more absorption features,
starting with 
Ti and then Ca. Abundances of these elements were then fixed at the fitted values, and further elements were fitted, etc.
Example fits are shown in Fig.~\ref{fig:fitfe}, where we plot the data together with the best-fitting model spectra and models where the Fe abundances were changed by $\pm0.3$ dex. Both models and data have been smoothed with a 5 pixels wide boxcar filter for the purpose of illustration.
The difference between the spectra of the very metal-poor clusters Fornax~3 and Fornax~5 and the significantly more metal-rich Fornax~4 is clearly seen.

Our individual abundance measurements are reported in Table~\ref{tab:rawabun}  together with the associated formal errors. When a wavelength \textit{range} is given in the first column, this indicates that several lines are present within that range but the overall abundance  was obtained from one fit. When instead a single wavelength is given, only one absorption line from the element is present, and the fit was carried out over a range of $\pm5$~\AA\, (roughly) centered on the wavelength of the feature. For windows wider than 10~\AA\ the continua of the model- and observed spectra were matched using a 3rd order spline with typically 5 knots, while a 2nd order polynomial was used for the smaller wavelength bins. 

In Table~\ref{tab:abun} we list the weighted average abundances per element for each cluster. In addition to the formal errors on the weighted means, we also list the r.m.s.\ bin-to-bin scatter of the individual abundance measurements when applicable. The latter may be viewed as a more realistic indicator of the actual measurement uncertainties, and may be turned into   errors on the mean abundances by taking into account the number of individual measurements ($N$), i.e., $\sigma = {\rm r.m.s} / \sqrt{N-1}$. In most cases, this yields errors  smaller than 0.1 dex, but nevertheless  somewhat larger than the formal errors based on the $\chi^2$ analysis. That the purely random errors do not fully account for the bin-to-bin scatter on the derived abundances can also be seen from the fact that some wavelength bins yield systematic higher or lower abundances than others. Indeed, the abundance measurements depend on other parameters, such as the atomic oscillator strengths of which some still remain quite uncertain (see e.g.\ discussion of Ti and Ba in Sect.~\ref{sec:letarte}) and hyperfine splitting.

\subsection{Comparison with Letarte et al.\ results for Fornax 3}
\label{sec:letarte}

\citet{Letarte2006} measured the abundances of several elements in \emph{individual} stars in Fornax~1, Fornax~2 and Fornax~3. In the following, we compare our integrated-light results with their data for three RGB stars in Fornax~3. 

\paragraph{Fe:}
L06 find an iron abundance of $\mathrm{[Fe/H]} =-2.4\pm0.1$. This is in excellent agreement with our integrated-light measurement of $\mathrm{[Fe/H]}=-2.33$. The r.m.s. scatter of 0.08 dex on our [Fe/H] measurements corresponds to an error on the mean of $\pm0.03$ dex.

\paragraph{$\alpha$-elements:}

For Ca, our measurement of $\mathrm{[Ca/Fe]}=+0.25$  agrees very well with the L06 result of [Ca/Fe]$=+0.22\pm0.02$. 
For Ti,  L06 measure $\mathrm{[\ion{Ti}{i} /Fe]}= -0.06\pm0.05$ and $\mathrm{[\ion{Ti}{ii}/Fe]} =0.10\pm0.02$. These values are both significantly lower than our measurement of $\mathrm{[Ti/Fe]} = +0.29$, which furthermore has a relatively small r.m.s.\ scatter of 0.11 dex with 9 individual measurements. As noted by L06,  the $\log gf$ values for Ti are somewhat uncertain, translating  into corresponding uncertainties on the derived abundances.
Comparing the $\log gf$ values for Ti used by L06 with those in the Kurucz line list, we find that the $\log gf$ values in the Kurucz list are generally  lower, with an average difference of about $\Delta$(Kurucz-L06) = $-0.15$ dex but with a range from $-0.8$ dex to $+0.15$ dex. Applying the average difference to the Ti abundances would bring our measurements in much better agreement with those of L06. However, this comparison is only indicative as the overlap between the lines we actually measure is limited.  L06 used a different UVES set-up, covering a redder wavelength range that does not include the range from 4200~\AA\ -- 4800~\AA\ where many of our Ti measurements are made.

Our [Mg/Fe] ratios are quite low compared to the other $\alpha$-elements, [Ca/Fe] and [Ti/Fe], for all three clusters. Here, the comparison with L06 is not very conclusive as L06 find a substantial spread in the $[$Mg/Fe$]$ values for their three stars  of $+0.19\pm0.09$, $+0.37\pm0.07$ and $-0.35\pm0.08$. This spread may be related to the anticorrelation of Al and Mg that is known to exist in some Milky Way GCs \citep{Shetrone1996,Gratton2001}. Neither we nor L06 measure Al, but it is interesting to note that low (solar-like) $[$Mg/Fe$]$  abundance ratios have also been found from integrated-light spectroscopy of M31 GCs \citep{Colucci2009}. \emph{It is therefore clear that caution should be exercised when using Mg as an indicator of the $[\alpha$/Fe$]$ ratio in the integrated light of GCs.}

\paragraph{Other elements:}

Other elements in common between our study and that of L06 are the Fe-peak elements Mn, Cr and the heavy elements Ba, Eu and Y. For Fornax~3 we find $\mathrm{[Cr/Fe]} =-0.17$ with an error of $\sim0.05$ dex based on the r.m.s. scatter. For their three stars in Fornax 3, L06 instead find $\mathrm{[Cr/Fe]} = -0.28\pm0.20$, $-0.30\pm0.20$ and $-0.50\pm0.15$.  The mean of the three L06 measurements is $-0.39\pm0.10$, formally a $\sim2\sigma$ difference with respect to our data. We note, however, that inclusion of low-mass stars below the adopted magnitude limit of our CMDs would tend to (slightly) decrease the Cr abundances (Sect.~\ref{sec:uncertainties} and Table~\ref{tab:systematics}).  Furthermore, L06 estimate a 0.21 dex model-dependent systematic uncertainty on their Cr abundance. We therefore conclude that our Cr abundance for Fornax~3 is consistent with the measurements of L06 within the uncertainties.

For Mn, we find a significantly sub-solar abundance ratio of $\mathrm{[Mn/Fe]}=-0.52$ while L06  find $\langle\mathrm{[Mn/Fe]}\rangle=+0.01$, in both cases with formal error bars less than 0.1 dex.  However, our [Mn/Fe] measurement is based on only a single wavelength bin in the blue, making it difficult to assess the real uncertainty. This wavelength range was not included by L06, while the 6014~\AA\ and 6022~\AA\ lines measured by L06 are too weak in our data to provide a reliable measurement. Indeed L06 only quote [Mn/Fe] for two or their three stars in Fornax~3, for one of three stars in Fornax~2 and for no stars in Fornax~1, attesting to the difficulty of these measurements. Our lower [Mn/Fe] abundance ratio is more similar to those seen in metal-poor Milky Way field stars and globular clusters and for field stars in the Fornax dSph, although  [Mn/Fe] ratios closer to Solar have also been derived from integrated-light measurements of LMC globular clusters \citep{Colucci2012}. We return to this point below.

Our abundances of Y, Ba and Eu ([Y,Ba,Eu/Fe] = $-$0.02, $+$0.81, $+$1.50) are somewhat higher compared to L06 ([Y,Ba,Eu/Fe] = $-$0.22, $+$0.22 and $+$0.90). 
For Ba in particular, the 0.6 dex difference is too large and too significant to be explained by measurement uncertainties. Our most discrepant measurement (relative to L06) comes from the 4934~\AA\ line. This line yields systematically higher Ba abundances than other lines not only for Fornax~3, but also for the two other clusters in our sample.
The $\log gf$ value for this line appears to be particularly uncertain: in the Kurucz line list it has $\log gf = -0.458$. L06 instead use $\log gf = -0.703$, while \citet{Colucci2009} use $\log gf = -0.157$. For the other Ba lines there are only minor differences between different sources. If we use the higher $\log gf$ value of \citet{Colucci2009} for the 4934~\AA\ line, our Ba abundance inferred from this line decreases accordingly and is then in better agreement with that inferred from the other Ba lines, and the average Ba abundances in  Table~\ref{tab:abun} decrease by $\sim0.15$ dex. If we instead use the smaller $\log gf$ value from L06, our Ba abundance increases correspondingly and the difference with respect to  L06   becomes even greater. 
The reason for this difference therefore remains unclear. However, some Galactic GCs are known to have a substantial internal spread in the heavy $n$-capture element abundances. In M15, the [Ba/Fe] ratio varies by 0.5--0.6 dex \citep{Sneden1997,Sobeck2011}, and is correlated with similar variations in the [Eu/Fe] ratio. These variations are seen also in the small sample of M15 stars measured by L06, and in the globular cluster M22 \citep{Marino2009}.  The uncertainties on our [Eu/Fe] abundances are large as the two lines we can measure are relatively weak and blended with several other lines, but it is worth noting that our [Eu/Fe] measurement does show a similar offset with respect to the L06 data as does [Ba/Fe]. This might suggest that the difference between our measurements and those of L06 are due to Fornax~3 having internally inhomogeneous $n$-capture element abundances, with the three stars measured by L06 happening to have less than average abundances of these elements. 
The light $n$-capture element Y does not appear to follow the same trends as Eu and Ba \citep{Sobeck2011}, and the difference between our [Y/Fe] ratio and that of L06 is indeed much smaller and not very significant given the $\sim0.10$ dex uncertainties  on both measurements. 
This scenario should, in principle, be verifiable by measuring a larger sample of individual stars in Fornax 3.
 
\subsection{Comparison with the spectrum of Arcturus}

\begin{table}
\begin{minipage}[t]{\columnwidth}
\caption{Abundances for Arcturus.
}
\label{tab:arcturus}
\renewcommand{\footnoterule}{}
\begin{tabular}{lccc} \hline\hline
               &  R11 & This work & r.m.s. \\ \hline
$[$Fe/H$]$    &  $-0.52\pm0.04$  & $-0.64$ & 0.04 \\
$[$Mg/Fe$]$ &  $+0.37\pm0.03$ & $+0.22$ & 0.06 \\
$[$Ca/Fe$]$ &  $+0.11\pm0.04$ & $+0.26^a$ & 0.17 \\
$[$Sc/Fe$]$  & $+0.21\pm0.04^b$ & $+0.11$ & 0.21 \\
$[$Ti/Fe$]$   & $+0.23\pm0.03^c$ & $+0.37$ & 0.31 \\
$[$Cr/Fe$]$  & $-0.05\pm0.04$ & $-0.04$ & 0.17 \\
\hline
\end{tabular}
\tablefoot{
\tablefoottext{a}{Rejecting $\lambda4878\AA$}
\tablefoottext{b}{Average of \ion{Sc}{i} and \ion{Sc}{ii}.}
\tablefoottext{c}{Average of \ion{Ti}{i} and \ion{Ti}{ii}.}
}
\end{minipage}
\end{table}

As an additional consistency check, we have applied our analysis to the spectrum of Arcturus \citep{Hinkle2005}. The physical parameters  of Arcturus are well established, and as a starting point for our analysis we used the stellar parameters determined by \citet[][hereafter R11]{Ramirez2011}: $T_{\rm eff} = 4286$ K and $\log g = 1.66$. These authors also find an iron abundance of [Fe/H] = $-0.52\pm0.04$, microturbulent velocity $v_t = 1.74$ km s$^{-1}$ and an absolute magnitude $M_V=-0.313\pm0.016$. For this $M_V$, our formula  yields $v_t = 1.725$ km s$^{-1}$, in close agreement with the value of R11.
We consider the step of deriving stellar parameters from integrated photometry a separate problem; this will be discussed below.

No special modifications to our procedure were required in order to model a single star -- we simply modelled the spectrum of Arcturus as a ``cluster'' of one star. 
However, because of the higher metallicity of Arcturus compared to the Fornax GCs, a much larger fraction of the spectrum is dominated by relatively strong absorption lines. In a proper analysis, this should be taken into account and strongly saturated lines should be avoided, but for consistency with the Fornax GC analysis we analysed the Arcturus spectrum using the exact same wavelength bins.
 However, due to the stronger lines in the Arcturus spectrum, we found that a simple spline fit  to the model/data ratio, as we did for the Fornax GCs, was significantly affected by several strong features in the observed spectrum which had no counterparts in the model. As a result, the continuum levels in the model spectra were noticeably lowered compared to the data, and the code would converge towards a mean iron abundance of [Fe/H]$\approx-0.70$, about 0.2 dex lower than the literature value. We therefore adopted a slight modification to the scaling procedure, in which we first identified ``continuum'' regions in the observed spectrum where the flux $f_\lambda > 0.98 f_{\rm max}$, where $f_{\rm max}$ was determined as a ``running maximum flux'' in 5 \AA\ bins. We then iteratively rejected pixels where the difference $(f_{\rm model} - f_{\rm data})$ exceeded 2$\times$ the standard deviation evaluated over all remaining pixels. With this modification to the scaling, we found that the continuum levels of the model- and observed spectra agreed well, and the
 average iron abundance increased to [Fe/H]=$-0.64$ with an r.m.s. scatter of $0.04$ dex. This is still about 0.1 dex lower than the R11 measurement, although lower values have been quoted in the literature \citep[e.g. \mbox{[Fe/H]}=$-0.68$;][]{Griffin1999}.

Apart from its overall lower metallicity, the abundance patterns in Arcturus differ from those of the Sun by being enhanced in the $\alpha$-elements \citep[e.g.][]{Peterson1993}. 
Abundances for elements in common between our work and R11 are compared in Table~\ref{tab:arcturus}.
Our measurements agree with those of R11 within 0.10--0.15 dex.
From the 4200\AA\ -- 4400\AA\ and 4400\AA\ -- 4600\AA\ bins we also derive a roughly Solar [C/Fe] ratio ([C/Fe]$\sim-0.05$). This is significantly lower than the R11 measurement of [C/Fe]$=+0.43\pm0.07$, while \citet{Peterson1993} quote [C/Fe]=0.0, similar to our measurement. The difference might be related to the use of individual \ion{C}{i} lines (R11) vs.\ the CH molecular bands \citep[this study and][]{Peterson1993}.

The transformation from broad-band photometry to stellar parameters is an important step in our procedure. Systematic errors in these transformations would translate to corresponding systematics in the abundance measurements. While we have used Kurucz transformations for solar-scaled abundances, using colour transformations based on $\alpha$-enhanced  model atmospheres only changes  $T_{\rm eff}$ by 10--15 K. This is in agreement with the findings of \citet{Cassisi2004}, who found that the $V\!-\!I$ colour is relatively insensitive to $\alpha$-enhancement (although other colours, like $B\!-\!V$ and $U\!-\!B$, are more strongly affected). However, colour-$T_{\rm eff}$ transformations differ significantly from one author to another. For example,
a comparison of the Kurucz $V\!-\!I$ vs. $T_{\rm eff}$ transformations with those of \citet{Worthey2011} shows differences of 50--70 K, with the Kurucz transformations tending to give cooler $T_{\rm eff}$ for giants and hotter $T_{\rm eff}$ for subgiants. Assuming $T_{\rm eff} = 4350$ K for  Arcturus (i.e., an increase of 64 K) and repeating the analysis, the iron abundance increases by 0.05 dex to [Fe/H] = $-0.59$. 
Overall, systematic errors in the colour-$T_{\rm eff}$ transformations have an effect that is quite similar to that of changing the correction for foreground extinction, which will be discussed in more detail below.

\subsection{Uncertainties on the abundance measurements}
\label{sec:uncertainties}

In Sect.~\ref{sec:results} we quantified the uncertainties on our abundance measurements due to formal  errors on individual wavelength bins and from the bin-to-bin scatter. However, in this type of analysis one must also consider a variety of systematic uncertainties, some of which we  discuss in the following.

\paragraph{Extinction.}

An increase in the extinction correction $E(V\!-\!I)$  leads to intrinsically bluer colours and higher $T_\mathrm{eff}$ for the stars. Overall, this has the effect of making most spectral lines in the model spectra weaker for a given metallicity, so the abundances inferred from the data increase to compensate. In Table~\ref{tab:systematics} we list the changes in the abundances for an increase of
$\Delta E(V\!-\!I) = +0.05$ mag. This corresponds roughly to the range of extinction values quoted in the literature for Fornax~4, while it probably exceeds the  uncertainties for Fornax~3 and 5 by a substantial margin.
For this $\Delta E(V\!-\!I)$,  the Fe abundances increase by $\Delta [$Fe/H$]\sim0.10$ dex. Abundance \emph{ratios} are much less strongly affected, and  most of the ratios we measure are hardly affected by uncertainties in the extinction correction at this level. The most strongly affected ratio is [Mg/Fe], which decreases by about 0.03 dex. Increasing the [Mg/Fe] ratios to bring them in better agreement with the other $\alpha$-element ratios would thus require a decreased $E(V\!-\!I)$, but even if we assume zero extinction towards each cluster, this would not make  the [Mg/Fe] ratios  similar to [Ca/Fe] or [Ti/Fe].

\paragraph{Microturbulence.}

An increase in the microturbulent velocity $v_t$ has the effect of de-saturating strong spectral lines, so a given observed line strength corresponds to a lower abundance. The magnitude of this effect depends on the strengths of the specific lines, with stronger (more saturated) lines  being more heavily affected. The sensitivity of our results to changes in $v_t$ was assessed by adding an offset of $\Delta v_t = 0.5$ km s$^{-1}$ to the assumed $v_t$ values and repeating the analysis. As can be seen in Table~\ref{tab:systematics}, the net effect of this (rather large) change is to decrease the Fe abundances by about 0.1 dex. Again, abundance ratios are generally less affected.
The most strongly affected abundance ratio is [Ba/Fe]; however, even large changes in $v_t$ would not be sufficient to bring our Ba abundance for Fornax~3 into agreement with that of L06.

\paragraph{Luminosity functions.}

The LFs are critical ingredients in our modelling of the integrated light as they determine the relative contributions from stars of different luminosities. Although Fig.~\ref{fig:lfs} shows good agreement between the model LFs and empirical  LFs for M92 within the range of $M_V$ values for which  data are available, subtle differences  might affect the abundance determinations at some level. To quantify this, 
we first replaced the Dotter et al. theoretical LFs with the empirical M92 LF for Fornax~3 and 5. For Fornax~4 we  used an empirical LF based on the globular cluster M12 \citep{Hargis2004}. We found that [Fe/H] values as well as individual abundance ratios changed by at most 0.010--0.015 dex. It thus appears that uncertainties due to the LF shape within this magnitude range are relatively minor.

At fainter magnitudes, dynamical effects become increasingly important. Our exclusion of stars fainter than $M_V=+9$ effectively corresponds to the extreme case in which no stars with $M\la0.4 M_\odot$ are present. To gauge the effect of fainter stars, we repeated the analysis for a limiting magnitude of $M_V=+13$, assuming a Salpeter IMF extending down to $0.1 M_\odot$. The last column in Table~\ref{tab:systematics} shows the effect of adopting this fainter magnitude limit.  The abundances  are indeed sensitive to the low-luminosity stars, with some  abundance ratios being at least as sensitive as the overall [Fe/H] abundance.  For realistic clusters, the assumption of a Salpeter IMF down to $0.1 M_\odot$ will probably \emph{overestimate} the contribution from the fainter stars, so we expect the changes to be smaller than the entries in the last columns of Table~\ref{tab:systematics}, probably on the order of a few times 0.01 mag for most ratios. Nevertheless, it is clearly  important for high-precision abundance analysis to pay attention to even the faintest stars, and the high mass-to-light ratios for Fornax~3 and 5 might indicate the presence of a relatively large number of low-mass stars \citep[][and Sect.~\ref{sec:mvir}]{Dubath1992}. 
With appropriate data, the contribution of dwarf stars to the integrated light might even be constrained as part of the analysis itself via features that are strongly sensitive to surface gravity \citep{Mieske2008,Conroy2012}.

\paragraph{Opacity distribution functions.} 

When computing the \texttt{ATLAS9} model atmospheres we  made use of pre-computed ODFs with solar-scaled abundance patterns. The ODFs are also available for $\alpha$-enhanced abundances ([$\alpha$/Fe] = $+0.4$). Computing the model atmospheres with the $\alpha$-enhanced ODFs and redoing the analysis, we found that this had a negligible effect on the results: [Fe/H] values changed by less than 0.005 dex, and most abundance ratios by less than 0.01 dex. The most extreme change was for the [Eu/Fe] ratio of Fornax~4, which decreased by 0.014 dex when using $\alpha$-enhanced ODFs. 

\paragraph{Stochastic sampling of the stellar IMF.}

It is well known that the finite number of stars in a cluster can cause its integrated colours to deviate strongly from those predicted for a continuously populated stellar mass function \citep[e.g.][]{Girardi1995,Bruzual2002,Fouesneau2010,Popescu2010a,Silva-Villa2011}.
We assessed the effect of random IMF sampling on our abundance determinations by bootstrapping. The original CMDs were resampled randomly with replacement and all steps of the analysis were then carried out again. This procedure was repeated 10 times, and the
uncertainties due to stochastic IMF sampling were then estimated as the standard deviation of the 10 resulting abundance determinations.

Due to the rather time consuming nature of this experiment we did not assess the effect on the full range of abundance ratios, but restricted the analysis to a few cases: [Fe/H], [Ti/Fe], [Ca/Fe], [Mg/Fe] and [Ba/Fe]. 
For [Fe/H] we found uncertainties of $\sigma_{\rm [Fe/H], stoc}$ = 0.02 dex, 0.03 dex and 0.05 dex for Fornax~3, Fornax~4 and Fornax~5, respectively. The scatter on the elemental abundance ratios was always less than on the [Fe/H] values.

Clearly, this procedure only gives an approximate description of the actual uncertainties. 
On the one hand, the stars in the HST CMDs are  \emph{actual} stars present in each cluster. Therefore, if our CMDs corresponded exactly to the rectangular regions covered by the drift-scanned spectra, stochastic fluctuations would play a less significant role than suggested by the bootstrapping experiment. However, even in this case it should be remembered that our CMD modelling is fully empirical only for the brighter stars. For fainter stars  we rely on model LFs to define the weight of each cmd-bin.
On the other hand, the CMDs  \emph{do not} correspond exactly to the parts of the clusters for which we have spectra. The central regions are too crowded to obtain reliable photometry, and  there is no photometry within $\sim2\arcsec$ of the cluster centres. 
From the profile fits of \citet{Mackey2003a}, about 10--15\% of the flux falls within the central region excluded from the photometry. 
Additionally, the observed CMDs include stars from regions of the clusters not covered by the spectra.  Roughly 2/3 of the stars in the CMDs fall within the slit area, while roughly 1/3 fall outside. 
Given the significant overlap, the true impact of stochasticity is likely smaller than estimated by our bootstrapping procedure.

\paragraph{Radial gradients.}

To test whether radial gradients in the integrated properties of the clusters contribute significantly to the uncertainties, we divided the spectral scans into an inner section (the central $\pm4$ pixels or $\pm1\farcs8$) and an outer section (the outermost 7 pixels on either side of the centre). The inner sections approximately span the radial range for which CMD data are missing, although the scans of course extend out to the half-light radius in both cases. We then analyzed each section separately. The differences were found to be very minor. For  [Fe/H] we found differences (outer-inner) of $0.001\pm0.009$ dex (Fornax 3), $-0.029\pm0.006$ dex (Fornax 4) and $-0.001\pm0.010$ dex (Fornax 5). Considering these very small differences, we did not look at other abundance ratios, but it appears safe to conclude that no strong radial trends are present.

\paragraph{Background subtraction}

The subtraction of background light from the spectral scans may also be affected by uncertainties due to stochastic sampling of the underlying field population.
However, as can be seen from Fig.~\ref{fig:F4sv}, the contrast between the GCs and background is quite high even for Fornax 4. 
In fact, with a central surface brightness of $\mu_V = 23.3$ mag arcsec$^{-2}$ for the Fornax dSph \citep{Irwin1995}, the background is everywhere dominated by the night sky rather than by the Fornax dSph itself. The total background signal, including night sky, constitutes  $\sim$3\%--6\% of the  counts in the science exposures, so the underlying field population in Fornax contributes only  marginally to the signal in each spectrum. This is corroborated by the analysis of \citet{Buonanno1999}, who find a ratio of cluster vs. field stars of about 4 at about 5$\arcsec$ from the centre of Fornax 4, corresponding to the end points of the slit in Fig.~\ref{fig:F4sv}. Integrating the fits of \citet{Mackey2003a} out to $r=5\arcsec$ then gives a \emph{number} ratio of cluster- to field stars of about 9.  Converting this to a luminosity ratio is less straight forward as the field- and cluster CMDs are quite different, due to the extended star formation history of the field. However, if we integrate the central surface brightness of the Fornax dSph over the central $5\farcs5$ of Fornax 4, and compare this with the apparent integrated cluster magnitude of $V=13.57$ \citep{Webbink1985} (adding 0.75 mag to find the magnitude within $r_{\rm eff}$) we find that the field contributes about 2.5\% to the signal. We therefore conclude that uncertainties due to the subtraction of the field star component from the spectral scans are very minor, even for Fornax 4.

\paragraph{Binning of the CMDs.}

As can be seen in Fig.~\ref{fig:cmds}, our adopted binning is sufficient to well sample all regions of the CMDs. We  verified that changes in the binning did not affect our results to any significant extent:  increasing the number of cmd-bins further only changed the [Fe/H] values by about 0.01 dex. In fact, a somewhat smaller number of bins would probably be sufficient; \citet{McWilliam2008} used only 27  bins in their modelling of the integrated spectrum of the globular cluster 47 Tuc.

\paragraph{Summary.}
From the above discussion, we conclude that the error sources discussed here all contribute by less than $\sim0.1$ dex to the uncertainties on our [Fe/H] measurements, and typically only by a few times 0.01 dex or less to abundance ratios. Nevertheless, the true uncertainties may well be dominated by effects that we cannot easily quantify here, such as non-LTE line formation, inclusion of scattering in the synthetic spectral calculations \citep{Cayrel2004}, 
accuracy of damping parameters and the atomic oscillator strengths and the quality and completeness of the line lists in general, as well as the simplifications involved in the use of classical model atmospheres.
Those limitations are not unique to our analysis, however, but affect most chemical abundance studies.

\subsection{Radial velocities}

The radial velocities of the clusters were determined with high accuracy as a byproduct of our analysis. The raw shifts required to match the data with the synthetic spectra were $-72.4$ km s$^{-1}$, $-59.4$ km s$^{-1}$ and $-72.7$ km s$^{-1}$ for Fornax~3, 4 and 5, respectively, with estimated uncertainties of  $\sim$0.2 km s$^{-1}$ for Fornax~3 and 5 and $<0.1$ km s$^{-1}$ for Fornax~4. After applying  heliocentric velocity corrections, this corresponds to radial velocities of 60.4 km s$^{-1}$, 47.2 km s$^{-1}$ and 60.6 km s$^{-1}$. Previously, \citet{Dubath1992} have reported radial velocities  of 58.5 km s$^{-1}$, 46.2 km s$^{-2}$ and 62.1 km s$^{-1}$ with errors of 1--2 km s$^{-1}$,  consistent with our measurements.

The average radial velocities for Fornax~1, 2 and 3 measured by \citet{Letarte2006} are 59.3, 63.8 and 62.7 km s$^{-1}$. However, these are based on measurements of three individual stars in each cluster and given the velocity dispersions of several km s$^{-1}$, these mean values must thus be considered uncertain by a few km s$^{-1}$. 

For comparison, the radial velocity of the Fornax dSph itself has been determined to be 53.3 km s$^{-1}$ and the velocity dispersion $\sigma=11.1$ km s$^{-1}$ \citep{Walker2006}.

\subsection{Velocity dispersions and dynamical masses}
\label{sec:mvir}

\begin{table}
\begin{minipage}[t]{\columnwidth}
\caption{Velocity dispersions, dynamical masses and mass-to-light ratios.
}
\label{tab:mvir}
\renewcommand{\footnoterule}{}
\begin{tabular}{lccccc} \hline\hline
                 & $\sigma_{\infty}$ (km s$^{-1}$) & $r_h$ (pc) & $M_{\rm vir}$ ($10^5 M_\odot$) & $M_V$ & $M/L_V$ \\ \hline
For 3 & $6.5\pm0.2$ & 7.2 & $5.2\pm0.4$ & $-8.2$ & $3.5\pm0.3$ \\
For 4 & $4.1\pm0.1$ & 4.9 & $1.4\pm0.1$ & $-7.4$ & $2.0\pm0.1$ \\
For 5 & $4.6\pm0.2$ & 8.5 & $3.2\pm0.2$ & $-7.5$ & $4.1\pm0.3$ \\
\hline
\end{tabular}
\end{minipage}
\end{table}

As another byproduct of the analysis, we briefly discuss the dynamics of the clusters.
The $\sigma_\mathrm{smooth}/\lambda$ ratios correspond to velocity broadenings of $c \sigma_\mathrm{smooth}/\lambda =  $ 8.1 km s$^{-1}$, 5.5 km s$^{-1}$ and 6.3 km s$^{-1}$. This may be compared with the (central) velocity dispersions for Fornax~3, 4 and 5, which have been measured to be 8.8 km s$^{-1}$, 5.1 km s$^{-1}$ and 7.0 km s$^{-1}$, with uncertainties of about 1 km s$^{-1}$ \citep{Dubath1992}. However, the instrumental broadening needs to be taken into account.
From a Gaussian fit to the sky emission lines at 5577~\AA\ and 5588/90~\AA\ we find a FWHM of 0.13 \AA, in excellent agreement with the resolving power $R$=40\,000 expected for a 1$\arcsec$ slit  according to the UVES manual. This corresponds to $\sigma_\mathrm{instr} \sim 0.06$ \AA\ or $\sim 3$ km s$^{-1}$. After subtracting this in quadrature from our measurements, we get $\sigma_\mathrm{vd} = 7.5$ km s$^{-1}$, 4.6 km s$^{-1}$ and 5.5 km s$^{-1}$ for Fornax~3, 4 and 5, respectively. Within the uncertainties, this agrees fairly well with the measurements of \citet{Dubath1992}. Our estimates tend to be systematically somewhat lower than theirs, but the errors on the Dubath measurements of the three clusters are probably not entirely independent as they are based on the widths of cross-correlation functions from the CORAVEL instrument, corrected for instrumental resolution with a set of reference stars common to all three clusters.
By integrating the projected velocity dispersion profiles of \citet{King1966} models with concentration parameters in the range $W_0 = 5-7$, we find that the central velocity dispersions are  5--10\% higher than the values measured within the half-light radius, an effect that works in the right sense but is not sufficiently strong to fully explain the difference between our measurements and those of Dubath et al. 
In any case, it is clear that the broadening of the spectral features in the GC spectra is dominated by the internal velocity dispersions for all three clusters.

From the velocity dispersions and the half-light radii, we can derive dynamical masses, $M_\mathrm{vir}$. We assume that
\begin{equation}
  M_\mathrm{vir} = 7.5 \frac{\sigma_{\infty}^2 r_h}{G}
  \label{eq:mvir}
\end{equation}
where $r_h$ is the three-dimensional half-mass radius, related to the projected two-dimensional half-light radius  $r_\mathrm{eff}$ as $r_h \approx \frac{4}{3} r_\mathrm{eff}$ \citep{Spitzer1987}. This assumes that mass follows light, which may not necessarily be the case if the clusters are mass segregated. Further, $\sigma_{\infty}$ denotes the global velocity dispersions. These are lower than our measured velocity dispersions $\sigma_\mathrm{vd}$, which only sample stars out to the half-light radii. We converted our velocity dispersion measurements within the rectangular regions covered by the UVES observations to global values by integrating over \citet{King1966} model profiles, as in \citet{Larsen2002b}. 

In Table~\ref{tab:mvir} we list the resulting global velocity dispersions, half-mass radii and virial masses as well as the $V$-band mass-to-light ratios ($M/L_V$).  The quoted errors only take into account the uncertainties on the velocity dispersions, determined from the scatter around the red lines in Fig.~\ref{fig:sigfit}. The $M_V$ magnitudes are based on the catalogue of \citet{Webbink1985}. Within the errors, our masses and $M/L_V$ ratios agree well with those of \citet{Dubath1992}. In particular, we confirm that Fornax~4 has a significantly lower $M/L_V$ ratio than the other, more metal-poor clusters. While this trend runs contrary to  predictions by standard SSP models, it is similar to that seen for a much larger sample of GCs in M31 \citep{Strader2009,Strader2011}. Indeed, the $M/L_V$ ratio of Fornax~4 is very similar to those of old M31 globular clusters of similar metallicity.  The $M/L_V$ ratios of Fornax~3 and 5 are very high, even by comparison with the most metal-poor clusters in M31 of which most have $M/L_V < 3$ \citep{Strader2011}. 
Dynamical effects will, in general, tend to \emph{lower} the observed mass-to-light ratios relative to canonical SSP model predictions, both via the preferential loss of low-mass stars \citep{Kruijssen2008}, and by leading to an underestimate of the true masses by dynamical measurements because the measured velocity dispersions and structural parameters will be most sensitive to the more massive stars, located preferentially near the centre \citep{Fleck2006,Miocchi2006}.
More exotic explanations include non-standard IMFs or even dark matter. Both top-heavy and bottom-heavy IMFs may in fact contribute to higher $M/L_V$ ratios - the latter due to the increased number of faint low-mass stars, and the former due to an increased number of dark remnants \citep{Dabringhausen2009}.  Binary stars could also lead to inflated velocity dispersions in integrated-light measurements. Common to all these potential explanations is that it is unclear why they should apply exclusively to the metal-poor Fornax GCs.

\section{Discussion}
\label{sec:discussion}

\subsection{Overall metallicities}

\begin{table}
\begin{minipage}[t]{\columnwidth}
\caption{Iron abundances, $[$Ca/Fe$]$ ratios and radial velocities for the five GCs in Fornax from high-dispersion spectroscopy.
}
\label{tab:fornaxgc}
\renewcommand{\footnoterule}{}
\begin{tabular}{lcccl} \hline\hline
                 & [Fe/H]                &   [Ca/Fe]  & $v_r$ (km s$^{-1}$) & Source \\ \hline
Fornax 1 & $-2.5\pm0.1$    & $+0.15\pm0.04$ & $59\pm1$ & L06 \\
Fornax 2 & $-2.1\pm0.1$    & $+0.20\pm0.03$ & $64\pm1$ &  L06\\
Fornax 3 & $-2.3\pm0.1$    & $+0.25\pm0.08$ & $60.4\pm0.2$ & This work\\
Fornax 4 & $-1.4\pm0.1$    & $+0.13\pm0.07$ & $47.2\pm0.1$ & This work \\
Fornax 5 & $-2.1\pm0.1$    & $+0.27\pm0.09$ & $60.6\pm0.2$ & This work \\
\hline
\end{tabular}
\tablefoot{For [Fe/H] the overall errors are set to $\pm0.1$ dex, including an estimate of systematic errors. For the remaining entries the errors are standard errors on the mean, based on the r.m.s.\ deviation of individual measurements.}
\end{minipage}
\end{table}

Combining the data from L06 and the present work, all five GCs in the Fornax dSph now have  chemical abundance measurements from high-dispersion spectroscopy. In Table~\ref{tab:fornaxgc} we summarize the [Fe/H] and [Ca/Fe] values, where the latter may be taken as a proxy for the [$\alpha$/Fe] ratio. For reference, we also include the radial velocity measurements, using our own measurements for Fornax~3, 4 and 5 and those of L06 for Fornax~1 and 2.
From this table it is now clear that the metallicity of Fornax~4 is indeed much higher (by almost a full dex) than the average of the other clusters. Because our measurements do not rely on any intermediate calibration steps of the metallicity scales, we believe this is a solid result. Moreover, any differences in the CMDs (such as the redder horizontal branch morphology of Fornax~4) are explicitly taken into account in our analysis.
We note that the CMD of Fornax~4 is, in fact, fairly well fit by a \citet{Dotter2007} isochrone with $\mathrm{[Fe/H]}=-1.5$ (Fig.~\ref{fig:cmds}). This is partly helped by the fact that we adopt a smaller $E(V\!-\!I)$ value than that of \citet{Buonanno1999},  making the RGB intrinsically redder and thus consistent with a higher metallicity. 

It is interesting to compare the metallicity distribution of the field stars in Fornax with the GC data. 
Field star metallicities have been measured by \citet{Battaglia2006} from \ion{Ca}{ii} IR triplet spectroscopy.
Comparing  their metallicity distribution for the field stars (their Figure 19), we see that  Fornax~1, 2, 3 and 5 all belong at the extreme metal-poor tail of this distribution. 
A more detailed analysis shows that a very large fraction, about 1/5--1/4,  of the most metal-poor stars in Fornax are found within the four most metal-poor GCs \citep{Larsen2012}.
The differences in chemistry are  correlated with the spatial locations of the clusters, with Fornax 4 being located close to the centre of the Fornax dSph while the other, more metal-poor clusters are found  further out, and in fact \citet{Strader2003} speculated that Fornax 4 may be the nucleus of the Fornax dSph. 
In this context it may be relevant that the radial velocity of Fornax 4 deviates by about 6 km s$^{-1}$ from the mean systemic velocity of the Fornax dSph.

We may also compare the metallicities of the Fornax GCs with those in the Milky Way. The McMaster catalog \citep{Harris1996} lists 11 clusters with $\mathrm{[Fe/H]}<-2$, a fraction of only 7\% of the Milky Way GC system. In striking contrast, 4 out of 5 GCs in Fornax have $\mathrm{[Fe/H]}<-2$.
In fact, the more ``metal-rich'' cluster, Fornax 4, has a metallicity  close to the peak of the metallicity distribution of Milky Way halo clusters, which is at $\mathrm{[Fe/H]}\approx-1.5$ \citep{Zinn1985}. It thus appears unlikely that a majority of the Milky Way GC system could have been accreted from Fornax-like dwarf galaxies.

\subsection{$\alpha$-elements}

The abundances of different $\alpha$-elements (O, Mg, Si, Ca, Ti) are generally found to correlate tightly with each other in field stars both in the Galaxy and in dSphs, with a tendency for the [Mg/Fe] ratio to be slightly higher than that of the other $\alpha$-elements in the dSphs \citep{Venn2004}. It is therefore interesting that we find a significantly \emph{lower} [Mg/Fe] ratio in the Fornax GCs compared to [Ca/Fe] and [Ti/Fe]. In fact, our [Mg/Fe] ratios  are consistent with being roughly Solar in all three GCs. A similarly low [Mg/Fe] ratio compared to other $\alpha$-elements has been found for the integrated light of GCs in M31 \citep{Colucci2009} and the LMC \citep{Colucci2012}. Colucci et al. found this to be accompanied by an elevated Al abundance. It is thus likely that we are detecting the signatures of the  abundance anomalies of light elements that are well-known in Galactic GCs, specifically the Mg-Al anticorrelation \citep{Gratton2001,Gratton2012}.
It is perhaps not highly surprising that these abundance anomalies are present in extragalactic GCs, given their ubiquity in Galactic GCs, but the demonstration that they are detectable in integrated light is a very important step towards establishing whether these anomalies are unique to \emph{ancient} GCs. Low-mass open clusters in the Milky Way do not display these patterns \citep{Pancino2010a}. However, it is unclear whether this implies fundamentally different origins for open and globular clusters, or simply reflects the large differences in the typical masses of objects belonging to either category that have been subject to detailed study. Any low-mass analogs of the surviving GCs have disrupted long ago, and are thus inaccessible to direct observations. A large fraction of the Galactic halo may indeed have formed originally in now-dispersed clusters \citep[e.g.][]{Kruijssen2009}, so the near-absence of the abundance anomalies in halo field stars \citep{Martell2010} is suggestive that these disrupted low-mass clusters were chemically more homogeneous internally than the surviving GCs. The remaining question, then, is whether  young analogues of globular clusters (at least in terms of their masses) that are forming in the present-day Universe share the same anomalous abundance patterns as their older cousins.
Although young clusters with masses $>10^5 M_\odot$ are rare or absent in the Local Group, they do exist in many external galaxies \citep{Larsen2000a,Larsen2009,Whitmore2003a}. If abundance anomalies are the norm in massive star clusters, a detailed analysis of the chemical composition of nearby young, massive clusters might reveal their presence also in such objects.

The remaining $\alpha$-elements that we could measure in the Fornax GCs (Ca, Ti) are clearly super-solar, but slightly less so than the typical enhancement of $\sim+0.4$ dex seen in Milky Way globular clusters and halo stars \citep{McWilliam1997}. In this respect, the Fornax GCs follow a trend seen also in the dSph field stars and in the Large Magellanic Cloud, which also tend to be  less $\alpha$-enhanced than Milky Way stars of similar metallicity \citep{Tolstoy2003,Venn2004,Johnson2006a,Kirby2011,Colucci2012}. For Fornax~3 and 5, both Ca and Ti are enhanced at the level of 0.25--0.30 dex, and there is a hint that Fornax 4 is even less $\alpha$-enhanced, with [Ca/Fe] and [Ti/Fe] ratios of $\sim0.12$ dex. This is interesting given the suggestion that Fornax~4 may have formed a few Gyr later than the other clusters \citep{Buonanno1999}. These authors also suggested that a non-standard chemical composition might be responsible for differences between the metallicities inferred from the RGB slope and other methods. Although we do find a difference, it is unclear whether a relatively modest difference of $\sim0.1$ dex between the [$\alpha$/Fe] ratios of Fornax~4 and the other clusters would be sufficient to explain this. 

\subsection{Fe-peak elements}

We measured several Fe-peak elements, namely: Sc, Cr, Mn, Fe and Cu. In spite of the common label assigned to these elements, it is not clear that they are all formed in a single nucleosynthetic channel and they do not all scale in a simple way with the Fe abundance. Indeed, we found both Mn and Cu to be significantly depleted with respect to Fe, with $\mathrm{[Mn/Fe]}\sim-0.3$ to $-0.5$ and $\mathrm{[Cu/Fe]}\sim-0.6$ to $<-1$, although it should be noted that our measurement of Cu is based on only a single line (5106~\AA ).
We found Solar or slightly sub-solar [Cr/Fe] values, while the [Sc/Fe] values are all consistent with Solar within the uncertainties.

It is  well established that Mn is depleted in metal-poor Milky Way halo stars and globular clusters, including $\omega$ Cen, at the level of $\mathrm{[Mn/Fe]}\sim-0.5$ \citep{Gratton1989,Nissen2000,Sobeck2006,Romano2011}, while at metallicities $\mathrm{[Fe/H]}>-1$ the [Mn/Fe] ratio gradually increases to Solar. A similar Mn depletion has been found in metal-poor field stars in dSph galaxies, including Fornax, and in the LMC \citep{Shetrone2003,Colucci2012,North2012}.
According to our measurements, the Fornax GCs are similar to metal-poor stars in the Fornax dSph and elsewhere in this respect. The usual interpretation of the low [Mn/Fe] ratios in metal-poor stars is that yields are strongly metallicity-dependent, although the actual formation site of Mn remains uncertain  \citep{Timmes1995,Shetrone2003}.
The higher Mn abundances derived for Fornax~3 (L06) and for a few LMC globular clusters \citep{Colucci2012} are then quite puzzling, and it would be desirable with more measurements in order to verify whether there is an offset between the [Mn/Fe] ratios in field- and globular cluster stars in some galaxies.
We note that the Mn features that are often measured by other authors (such as the \ion{Mn}{i} triplet near 6020 \AA) are mostly too weak in the Fornax GCs and our Mn abundances mainly rely on the less commonly used \ion{Mn}{i} lines near 4770 \AA . These do not have hyperfine splitting included in the line list we are using. However, the Kurucz website now includes line lists with hyperfine splitting for many more transitions, including these lines, and we have verified that the Mn abundances change very little (decreasing by 0.01--0.05 dex, with the largest change for Fornax 4) if we use the new line lists.

The behaviour of Cu is fairly similar to that of Mn. At $\mathrm{[Fe/H]}\sim-2$, [Cu/Fe] ratios are typically in the range $-0.9$ to $-0.5$ for both field and GC halo stars, with an increase in the [Cu/Fe] ratio at higher [Fe/H] \citep{Sneden1991a,Simmerer2003}. In this respect, our measurements of Fornax 3 and 5 are similar to those of Milky Way stars with similar metallicity. However,  Fornax 4 with its higher overall metallicity falls significantly below the trend found for Milky Way stars. At $\mathrm{[Fe/H]}=-1.4$, the typical Cu abundance in Milky Way stars is $\mathrm{[Cu/Fe]}\sim-0.5$, albeit with a 0.1-0.2 dex scatter. Our lower [Cu/Fe] ratio of $-0.8$ for Fornax~4 mimics the patterns observed in the LMC and dSph field stars, which also tend to fall below the Milky Way [Cu/Fe] vs. [Fe/H] relation \citep{Shetrone2003,Shetrone2004,Johnson2006a}. 

The Sc abundance generally tends to follow Fe, both in metal-poor stars in the Milky Way \citep{Gratton1991},  in the LMC \citep{Johnson2006a} and in dSphs \citep{Shetrone2003}. However, \citet{Nissen2000} find [Sc/Fe] decreasing from $\mathrm{[Sc/Fe]}\sim+0.2$ at $\mathrm{[Fe/H]}<-1$ to [Sc/Fe]$\sim0$ at solar metallicity, and suggest that Sc (intermediate in atomic number between Ca and Ti) behaves like an $\alpha$-element. Our data suggest no significant enhancement of the [Sc/Fe] ratio in the Fornax GCs.

Concerning Cr, \citet{Johnson2006a} find solar to slightly sub-solar [Cr/Fe] ratios for LMC GCs, and stars in the Milky Way appear to have [Cr/Fe] ratios very close to 0 over a wide range of metallicity \citep{Gratton1991}.  This is again similar to our results for the Fornax GCs.

\subsection{Neutron-capture elements}

We measured the abundances of  the three neutron-capture elements Ba, Y and Eu. Of these, Eu is considered a pure $r$-process element, while Ba and Y can also be produced via the $s$-process. As such, the abundance of Eu relative to the other elements is an indicator of the extent to which the nucleosynthesis was dominated by the $r$-process prior to the formation of the GCs. Since the site of the $r$-process is usually assumed to be massive stars, a strong $r$-process contribution is then taken as an indication that the stars in question formed on a short time scale.

In the Milky Way halo, the neutron capture elements are characterized by a very large abundance spread relative to Fe, usually taken as evidence of inhomogeneous chemical evolution. Notably, this spread is also seen internally in some GCs (Sect.~\ref{sec:letarte}).
At metallicities $\mathrm{[Fe/H]}<-2$, the Ba abundance spans a wide range of $-1<\mathrm{[Ba/Fe]}<1$, with a few even more extreme outliers \citep{McWilliam1997,Francois2007}. Such a spread is also seen in dSphs \citep{Venn2004}. 

We found a rather large spread in the [Ba/Fe] ratios of the Fornax GCs, similar to that seen in the Galactic halo. 
Eu is also displaying a substantial spread, but our measurements of this element are quite uncertain. Nevertheless, the Eu abundances in all three clusters appear strongly enhanced relative to both Fe and Ba with an average $\mathrm{[Ba/Eu]}\sim-0.7$. This
[Ba/Eu] ratio approaches that of the pure $r$-process \citep[$\mathrm{[Ba/Eu]}=-0.8$;][]{McWilliam1997}, indicating a dominant contribution to neutron capture synthesis by the $r$-process.
This is again similar to the patterns observed in metal-poor stars in the Milky Way, as well as in the LMC and Local Group dSphs \citep{Venn2004,Johnson2006a,Colucci2012}.
The [Ba/Eu] ratio of Fornax~4 is not significantly different from that of the other clusters -- in fact, Fornax~4 formally has the lowest [Ba/Eu] ratio, and thus the strongest $r$-process signature, of the three. This may be somewhat contrary to what we might expect if Fornax~4 were several Gyr younger than the other clusters, but the large uncertainties on the Eu abundances should be kept in mind here. They are based on measurements of two lines at 4205~\AA\ and 4436~\AA, both of which are rather complex blends. 

The [Ba/Y] ratio is interesting as it measures the ratio of heavy vs.\ light neutron capture elements, and is less uncertain than the ratios involving Eu. We found a mean $\mathrm{[Ba/Y]}\sim+0.66$, very similar to the mean [Ba/Y] ratio for field stars in dSphs and LMC \citep{Shetrone2003,Venn2004,Colucci2012}, but significantly more positive than for metal-poor Milky Way stars  \citep{Francois2007}.

\section{Conclusions}
\label{sec:conclusions}

We have carried out a detailed analysis of metallicities and abundance patterns for the three globular clusters Fornax~3, 4 and 5 in the Fornax dwarf spheroidal galaxy. Our analysis is based on the integrated light of the clusters, using drift-scan observations with the UVES spectrograph on the ESO Very Large Telescope, and a combination of Hubble Space Telescope observations of the colour-magnitude diagrams and theoretical models to characterize the stellar contents of the clusters. We have introduced and described the method we have developed for the analysis of such data, and carried out several tests to assess various systematic and random errors.
Our main results are as follows:

\begin{itemize}
\item Combining our data with measurements of individual stars in Fornax 1 and 2 \citep{Letarte2006}, it is clear that the Fornax GCs  fall into two distinct groups: Fornax 1, 2, 3 and 5 have very low metallicities in the range $-2.5 <\mathrm{[Fe/H]}<-2$, well below the average metallicity of Milky Way halo GCs, while Fornax~4 has $\mathrm{[Fe/H]}\sim-1.4$. Taking into account both random and systematic errors, we estimate the uncertainties on the iron abundances to be $\sim0.1$ dex.
It is thus firmly established that Fornax~4 is significantly more metal-rich than the other four clusters.
\item Our integrated-light measurements of a wide range of elemental abundances in the Fornax GCs are mostly consistent with observations of individual metal-poor stars in dSphs, with the notable exception of [Mg/Fe]. Specifically, we find a more moderate $\alpha$-enhancement (based on [Ca/Fe] and [Ti/Fe]) than typical of metal-poor stars in the Milky Way, especially for Fornax 4. We also find a strong $r$-process signature, even in Fornax~4 despite the younger age suggested by some studies and the low [$\alpha$/Fe] ratio.
\item In all three clusters, the [Mg/Fe] ratio is significantly lower than [Ca/Fe] and [Ti/Fe]. This is likely a signature of the Mg-Al abundance anti-correlation that is well-known from some Milky Way globular clusters. While it is very encouraging that such abundance anomalies may be detectable in integrated light, this also implies that [Mg/Fe] may be a poor proxy for the overall [$\alpha$/Fe] ratio, at least for globular clusters.
\item We confirm that Fornax~4 has a significantly \emph{lower} mass-to-light ratio than the more metal-poor clusters Fornax~3 and Fornax~5. This trend, while opposite to predictions by SSP models, is similar to that found for globular clusters in M31 \citep{Strader2009,Strader2011}.
\end{itemize}

Having the ability to carry out detailed abundance analysis from integrated light means that this type of analysis can now be extended to galaxies well beyond the Local Group. We believe that, for the time being, it is still risky to rely on purely theoretical models for the underlying  H-R diagrams of the clusters, particularly at young ages where such models are still quite uncertain. However, colour-magnitude diagrams are available for several young massive star clusters in galaxies within distances of several Mpc \citep{Larsen2011}, allowing detailed integrated-light abundance analysis for such clusters to be put on a solid footing. In the slightly more distant future, the combination of high-resolution imaging and high-dispersion spectroscopy on 30-40 m class telescopes will be a very powerful tool for detailed analysis of the chemical composition of star clusters in a significant volume of the Local Universe.

\begin{acknowledgements}
When we discussed the  M33 globular cluster spectra published in \citet{Larsen2002b}, John Huchra posed the question: \textit{What are the plans to say anything about the line strengths and abundances from the HIRES spectra?} The analysis presented here would not have been computationally feasible in its current form with the hardware available to us then, but it is appropriate to acknowledge John's role in stimulating our thoughts on this issue. We further thank Ruth Peterson for discussions and advice,  C.\ E.\ Corsi for kindly providing the photometry of Fornax~3 and 5, and
the referee for a very careful reading of the manuscript and many helpful suggestions.
This research has made use of NASA's Astrophysics Data System Bibliographic Services.
SSL acknowledges support by an NWO/VIDI grant and JPB acknowledges NSF grant AST-1109878.
\end{acknowledgements}

\bibliographystyle{aa}
\bibliography{libmen.bib}

\onecolumn

\end{document}